\documentclass[a4paper,11pt]{article}
\pdfoutput=1
\usepackage{jheppub}

\usepackage[utf8]{inputenc}
\usepackage{graphicx}
\usepackage{grffile}
\usepackage{dcolumn}
\usepackage{makecell}
\usepackage{amssymb}
\usepackage{amsmath}
\usepackage{bm}
\usepackage{subfig}
\usepackage{natbib}

\usepackage{eso-pic}
\usepackage[colorlinks=true,
urlcolor=blue,
linkcolor=blue,
citecolor=blue,
linktocpage=true,
pdfproducer=medialab,
pdfa=true,
anchorcolor=blue]{hyperref}
\usepackage{tabularx,array}
\usepackage[percent]{overpic}

\newcolumntype{C}{>{\centering\arraybackslash}X}
\usepackage{multirow}
\usepackage{float}
\usepackage{xcolor}
\usepackage{booktabs}

\renewcommand{\thesection}{\arabic{section}}

\newcommand{\eelogo}{%
  \AddToShipoutPictureBG*{%
    \put(\LenToUnit{\dimexpr 1in+\oddsidemargin\relax},%
         \LenToUnit{\dimexpr\paperheight+\voffset-15mm\relax}
         ){%
      \begin{minipage}[c]{15mm}%
        \includegraphics[width=\textwidth]{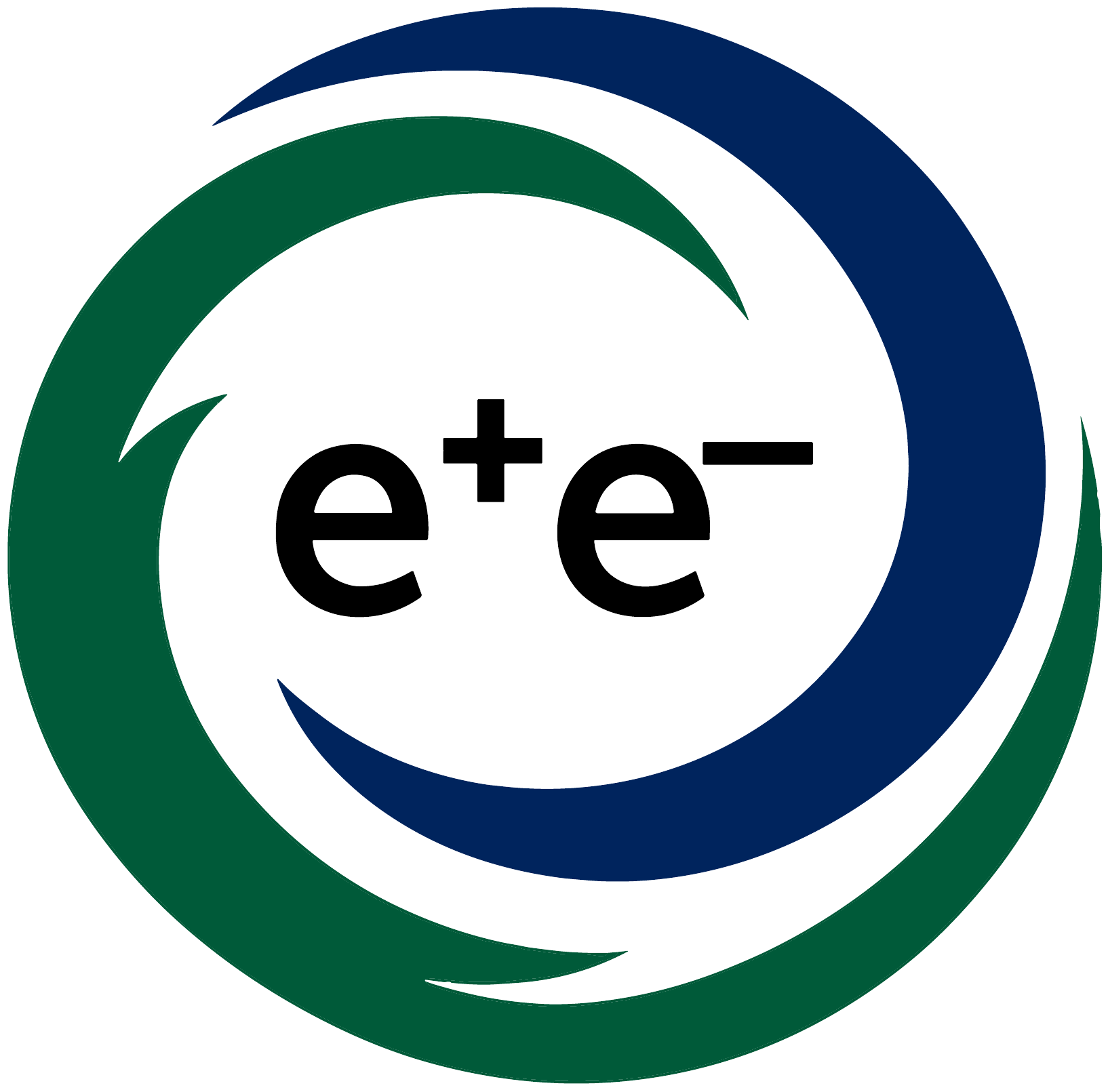}%
      \end{minipage}%
      \hspace{4mm}%
      \raisebox{-1mm}[0pt][0pt]{%
        \begin{minipage}[c]{0.44\textwidth}%
          \normalsize
          Electron--Positron Alliance%
        \end{minipage}}}}%
}
\input belle2sym.tex

\preprint{}
\makeatletter
\gdef\@fpheader{}
\makeatother

\begin{document}

\title{First measurement of the one-point charge correlator in $e^+e^-$ collisions at $\sqrt{s} = 91.2$~GeV with DELPHI Open Data}

\author[a]{Jingyu Zhang,}
\author[a]{Yi Chen,}
\author[b]{Kyle Lee,}
\author[c]{Ian Moult,}
\author[d]{Cristian Baldenegro,}
\author[d]{Hannah Bossi,}
\author[d]{Yen-Jie Lee}

\affiliation[a]{Department of Physics and Astronomy, Vanderbilt University, Nashville, Tennessee, USA}
\affiliation[b]{High Energy Physics Division, Argonne National Laboratory, Lemont, IL, USA}
\affiliation[c]{Department of Physics, Yale University, New Haven, CT 06511, USA}
\affiliation[d]{Laboratory for Nuclear Science, Massachusetts Institute of Technology, Cambridge, Massachusetts, USA}

\emailAdd{jingyu.zhang@cern.ch}
\arxivnumber{2608.15357}
\eelogo

\abstract{
The chiral structure of the $Z$ couplings imprints a parity-odd flow of electric charge on hadronic $Z$ decays. The related forward-backward asymmetries, a key set of observables in the electroweak precision program, were measured at LEP and SLC using the jet charge. The one-point charge correlator offers a complementary route, measuring the hadronic charge flow directly as a function of polar angle relative to the incoming electron-beam axis, without reference to jets or a reconstructed quark direction, following the formalism developed in a companion paper. We report its first measurement, using $61~\mathrm{pb}^{-1}$ of archival DELPHI Open Data recorded at $\sqrt{s} = 91.2$~GeV in 1994 and 1995. Detector effects are corrected in two stages. The first is derived from fully simulated samples, and the second bounds the residual charge-misreconstruction difference between data and simulation using a measurement in $e^+e^-\to\tau^+\tau^-$ events. The measured charge correlator exhibits the characteristic parity-odd $\sin(2\theta)$ modulation and agrees with the \textsc{PYTHIA}~8.3 prediction. The measurement demonstrates that the parity-odd charge flow of hadronic $Z$ decays is directly accessible in data, differentially in polar angle, and the paper discusses strategies for controlling the associated detector effects, paving the way for a new program of charge-flux measurements, both in archival $e^+e^-$ data and at future colliders.} 

\keywords{
Electron-positron collisions,
charge correlators,
electroweak physics,
DELPHI Open Data
}

\maketitle

\section{Introduction}
\label{sec:intro}

The large samples of $Z$ bosons produced at the LEP and SLC detectors played a key role in the establishment of the electroweak sector of the Standard Model. Precision measurements of the $Z$ boson total width, its couplings to leptons and quarks, and the electroweak mixing angle enabled the first tests of electroweak symmetry breaking at the quantum level, famously enabling an indirect determination of the top quark mass and constraining the Higgs boson mass range~\cite{ALEPH:2005ab}. With the successful running of the Tevatron and the LHC, precision measurements of the W, top, and Higgs masses, as well as Higgs couplings, now overconstrain global electroweak fits, providing some of the best tests of the structure of the Standard Model and some of the most stringent constraints on physics beyond the Standard Model~\cite{Baak:2014ora, Haller:2018nnx, Fischer:2026bka}. The $Z$-pole inputs to these fits, however, remain those of LEP and SLC.

Among the key $Z$-pole observables in the electroweak precision program are the forward--backward asymmetries, which are sensitive to the chiral structure of the $Z$ couplings and provide the primary determination of the electroweak mixing angle. These have been measured for both leptons and quarks, and as a function of flavor. The forward-backward asymmetry for b-quarks exhibits a longstanding tension with the Standard Model, and indeed is the largest tension in precision electroweak fits. This has generated significant interest, since many models of physics beyond the Standard Model couple most strongly to the third generation and can naturally explain such a modification relative to the Standard Model. Well-studied examples include vector-like b-quarks and composite Higgs models~\cite{Choudhury:2001hs, Agashe:2006at}.

Measurements of forward-backward asymmetries are straightforward for leptons, since their kinematics can be directly measured in the detector. On the other hand, measurements of hadronic asymmetries are highly non-trivial: the angular structure of the coupling of the Z boson to quarks must be reconstructed from hadrons in the detector. Hadronic forward--backward asymmetries were measured extensively at LEP and SLC: inclusively on all hadrons \cite{OPAL:1997tsq,OPAL:1992jsm,ALEPH:1991fba,ALEPH:1996qlh,ALEPH:1998pmr,L3:1991gfs,L3:1998jet,DELPHI:1991mqi}, on $b$- and $c$-enriched samples \cite{OPAL:1993wua,ALEPH:2001mdb,L3:1992fsb,DELPHI:2004wvq,OPAL:2002ddm,L3:2000vgx,SLD:2005gev}, and on $s$-enriched samples \cite{DELPHI:1994aml,SLD:2000jop,DELPHI:1999mkl}, with the combined electroweak results summarized in Refs.~\cite{ALEPH:2005ab,ALEPH:2010aa}. These analyses were built on the jet charge \cite{Field:1977fa}, the momentum-weighted sum of track charges within a jet or thrust hemisphere, often combined with leptons from heavy-flavor decays and secondary-vertex information. The interpretation of these measurements is, however, limited by the observables themselves. The jet charge involves jet and hemisphere definitions and their association with a primary quark, and is not infrared and collinear safe \cite{Catani:1999nf,Weinzierl:2006yt}, so its quantum chromodynamics (QCD) corrections cannot be computed fully perturbatively, and the residual model dependence is difficult to quantify from first principles \cite{dEnterria:2018jsx}. This has made it challenging to confront these measurements with improved theory, and as such the longstanding tension in global electroweak fits has persisted for decades.

Revisiting this tension therefore calls for an observable whose relation to the underlying couplings is not mediated by such an association. One would instead like to extract the coupling structure of the $Z$ boson directly from asymptotic fluxes. This direct study of correlations in asymptotic fluxes, and of the states that produce them, was re-initiated in Ref.~\cite{Hofman:2008ar}, and has been most actively pursued in the context of energy correlators~\cite{Moult:2025nhu}, which have proven both experimentally measurable across a variety of collider systems and theoretically clean, providing a productive interface between theory and experiment.

Forward-backward asymmetries can also be formulated in terms of correlations in flux, but instead of energy flux, they appear in fluxes of charges, such as electromagnetic charge. The application to hadronic forward--backward asymmetries is developed in a companion paper~\cite{Zhang:2026emt}. In particular, the forward-backward asymmetry can be captured by the one-point correlator of charge flux. The one-point charge correlator offers a complementary approach to previous studies of forward-backward asymmetries. It requires no jets, hemispheres, or primary-quark assignment, and measures the charge flux differentially in polar angle, thereby providing an independent path to the hadronic asymmetries. However, the one-point correlator of charge flux has never been experimentally measured.

Recent analyses have illustrated the potential of archival data for QCD and precision physics \cite{Electron-PositronAlliance:2019cpi,Chen:2021uws, Chen:2023njr, Electron-PositronAlliance:2025fhk, Electron-PositronAlliance:2025hze}. While these measurements have focused on event shape studies of QCD, an unexplored potential of this dataset, being at the $Z$ pole, is for precision electroweak physics. This offers an exciting opportunity to re-perform precision electroweak measurements using modern observables, theory, and data-analysis techniques, and to resolve or confirm the longstanding tensions in electroweak fits. 

This work presents the first measurement of the one-point charge correlator at the $Z$ pole, utilizing DELPHI Open Data~\cite{DELPHI:2024opendata} at LEP-1 at $\sqrt{s} = 91.2$~GeV during 1994 and 1995. The goal of the present measurement is to establish the observable experimentally. A precision extraction of the hadronic asymmetries is left to future work. An outline of this paper is as follows. Section~\ref{sec:observable} defines the one-point charge correlator and its binned estimator. Section~\ref{sec:data_samples} describes the detector, the data and simulated samples, and the track and event selections. A combined simulation-based and data-driven correction strategy is then applied to the raw data. Section~\ref{sec:mc_correction} presents the simulation-based correction, derived and applied separately for positive and negative tracks. Section~\ref{sec:cp_correction} discusses the data-driven charge calibration strategy, and the charge-misreconstruction rate is measured in a control region of $e^+e^- \to \tau^+\tau^-$ events in Section~\ref{sec:data_driven}. Sections~\ref{sec:systematics} and~\ref{sec:results} present the statistical and systematic uncertainties and the final results, respectively. Section~\ref{sec:summary} summarizes the measurement and its outlook. A detailed account of the analysis, including additional data--simulation comparisons, control-sample studies, and cross-checks of the analysis choices, is given in a companion analysis note~\cite{Zhang:2026esd}.


\section{Definition of the observable}
\label{sec:observable}

At the $Z$ pole, the Born-level differential cross section for $e^+e^- \to q\bar{q}$ takes the form
\begin{equation}
\frac{\mathrm{d}\sigma}{\mathrm{d}\cos\theta} \;=\;
\frac{3\,\sigma_{q\bar{q}}}{8}\left(1 + \cos^2\theta\right)
\;+\; \sigma_{q\bar{q}}\,A_{\rm FB}^{q}\,\cos\theta ,
\label{eq:diffxsec}
\end{equation}
where $\theta$ is the polar angle of the outgoing quark with respect to the $e^-$ beam direction and $\sigma_{q\bar{q}}$ is the total $Z\to q\bar{q}$ cross section. The two angular terms have distinct symmetry under $\theta \leftrightarrow \pi - \theta$. The $(1+\cos^2\theta)$ component is parity-even and dominates the rate, while the parity-odd $\cos\theta$ component carries the entirety of the parity-violating information, with coefficient the forward--backward asymmetry 
\begin{equation}
A_{\rm FB}^{q} = \frac{\sigma_{\rm F} - \sigma_{\rm B}}{\sigma_{\rm F} + \sigma_{\rm B}},
\end{equation}
where $\sigma_{\rm F}$ and $\sigma_{\rm B}$ are the cross sections for the quark to be produced in the forward and backward hemispheres.

The quark direction appearing in Eq.~\eqref{eq:diffxsec} is not itself an observable: confinement means that only the final state hadrons reach the detector. Let $j^{\mu}_{\rm em}$ denote the electromagnetic current carried by the hadrons. Following Ref.~\cite{Hofman:2008ar}, the associated charge flux operator is
\begin{equation}
    \mathcal{Q}(\vec{n})
    \;=\;
    \lim_{r\to\infty} r^{2}\!\int_{0}^{\infty}\!\mathrm{d}t\;
    n_i\, j^{\,i}_{\rm em}\!\left(t,\, r\vec{n}\right),
    \label{eq:fluxop}
\end{equation}
where $\vec{n}$ is a unit vector. Equation~\eqref{eq:fluxop} measures the net electric charge deposited per unit solid angle in the direction $\vec{n}$. It is the charge analog of the energy flux operator obtained by replacing $j^{\,i}_{\rm em}$ with the stress tensor. The one-point charge correlator is the expectation value of this operator, $\langle Q(\vec{n})\rangle$, in the state created by the electroweak current. It is accessible experimentally by acting with the charge flux operator on the final state $\left|X\right\rangle$ of hadrons that reach the detector,
\begin{equation}
    \mathcal{Q}(\vec{n})\left|X\right\rangle
    \;=\;
    \sum_{i \in X} q_i\, \delta\!\left(\Omega - \Omega_i\right)\left|X\right\rangle ,
    \label{eq:fluxstate}
\end{equation}
where the sum runs over the particles of the final state $X$ with electric charges $q_i$, and $\Omega$ and $\Omega_i$ denote the solid angles of $\vec{n}$ and of particle $i$. The correlator is therefore a normalized charge-weighted cross section, in which each final state contributes the charges of its particles distributed according to their directions. 

For unpolarized beams, the initial state is invariant under rotations about the beam axis, so $\langle \mathcal{Q}(\vec{n}) \rangle$ depends only on the polar angle $\theta$, and $CP$ invariance of inclusive hadronic $Z$ decays requires it to be exactly odd under $\theta \to \pi - \theta$. The parity-even $(1+\cos^{2}\theta)$ term that dominates Eq.~\eqref{eq:diffxsec} therefore contributes to the total charged-particle density but not to the charge flux. The charge flux inherits the $\cos\theta$ structure of the Born-level distribution, as formulated in the companion paper. The measurement is performed in polar angle rather than in solid angle. The corresponding density $\mathcal{Q}(\theta)$ is the charge flux per unit polar angle, obtained from $\langle \mathcal{Q}(\vec{n}) \rangle$ by integrating over the azimuth angle and including the $\sin\theta$ Jacobian relating solid angle to polar angle,
\begin{equation}
    \mathcal{Q}(\theta)
    \;=\;
    2\pi \sin\theta\,\left\langle \mathcal{Q}(\vec{n}) \right\rangle
    \;\propto\;
    \sin\theta\,\cos\theta
    \;\propto\;
    \sin(2\theta).
    \label{eq:qtheta_from_flux}
\end{equation}
A $\sin(2\theta)$ modulation of the measured distribution therefore is expected from this Jacobian.

Written out as a normalized charge-weighted cross section, $Q(\theta)$ is a direct sum over reconstructed charges,
\begin{equation}
    \mathcal Q(\theta)
    =
    \left\langle \sum_{i\in X} q_i\,\delta(\theta-\theta_i) \right\rangle
    =
    \frac{1}{N_{\rm evt}}
    \sum_{e=1}^{N_{\rm evt}}
    \sum_{i\in e} q_i\,\delta(\theta-\theta_i), 
    \label{eq:qdef}
\end{equation}
where $N_{\rm evt}$ denotes the total number of events. 
The measurement evaluates this continuous distribution in finite angular bins. For a bin $a$ of width $\Delta\theta_a$, let
\begin{equation}
    n_a^{\pm}
    \equiv
    \frac{N_a^{\pm}}{N_{\rm evt}\,\Delta\theta_a}
    \label{eq:charged_densities}
\end{equation}
denote the positive- and negative-track densities, where $N_a^{\pm}$ are the corresponding total track yields. The measured value in the bin is the bin average of Eq.~\eqref{eq:qdef},
\begin{equation}
    \mathcal Q_a
    \equiv
    \frac{1}{\Delta\theta_a}
    \int_{\mathrm{bin}\,a} \mathcal Q(\theta)\,\mathrm d\theta
    = n_a^+ - n_a^-,
    \label{eq:qNN}
\end{equation}
and the corresponding total charged-particle density is
\begin{equation}
    N_{{\rm tot},a} \equiv n_a^+ + n_a^-.
    \label{eq:ntot}
\end{equation}
Thus, $\mathcal Q_a$ is the experimentally measured, binned representation of the charge correlator $\mathcal Q(\theta)$. In the following, $\mathcal Q(\theta)$ and $N_{\rm tot}(\theta)$ are used for these binned densities, plotted at the angular-bin centers. Bin widths are expressed in radians.

Hadrons in the forward hemisphere, $\theta < 90^\circ$, carry on average a net charge correlated with that of the quark, and those in the backward hemisphere with that of the antiquark, since the quark is produced preferentially in the forward direction. The charge-symmetric part of fragmentation instead produces charge-balanced pairs that are nearby in angle and compensate locally. The size of the modulation is proportional to the flavor-weighted combination $A_{\rm FB}=\sum_q R_{q}\,Q_q\,A_{\rm FB}^{q}$, where $R_q = \Gamma_{q\bar{q}}/\Gamma_{\rm had}$ is the hadronic branching fraction of flavor $q$, $Q_q$ its electric charge, and $A_{\rm FB}^{q}$ its forward--backward asymmetry.

\section{Detector, datasets, simulations, and selections}
\label{sec:data_samples}

DELPHI was one of the four large general-purpose detectors at LEP and collected data from 1989 to 2000. A comprehensive description of its design and performance can be found in Refs.~\cite{DELPHI:1990cdc, DELPHI:1995dsm}. It comprises over 20 sub-detectors arranged in a cylindrical geometry around the interaction point, with charged-particle tracking performed in a 1.2~T solenoidal magnetic field parallel to the beam axis. The tracking system, the component most relevant to this measurement, consists of the silicon Microvertex Detector, the Inner Detector drift chamber, the Time Projection Chamber (TPC), and the Outer Detector, complemented in the forward regions by the FCA and FCB planar drift chambers. The TPC, the main tracking device, reconstructs tracks with up to 16 space points between polar angles of $39^\circ$ and $141^\circ$, and with at least four space points down to $21^\circ$ and up to $159^\circ$. Importantly, the apparatus is symmetric in the azimuthal angle $\phi$ and, to good approximation, forward--backward symmetric under $\theta \to 180^\circ - \theta$. The detector geometry therefore does not by itself induce a parity-odd charge signal, and instrumental contributions enter only through residual charge- and $\theta$-dependent responses, which are the subject of the corrections described in Sections~\ref{sec:mc_correction} and~\ref{sec:data_driven}.

This analysis uses data collected by DELPHI at a center-of-mass energy of $\sqrt{s} = 91.2$~GeV during 1994 and 1995~\cite{DELPHI:OpenData:short94_c2, DELPHI:OpenData:short95_d2}. The 1994 data correspond to an integrated luminosity of 46~pb$^{-1}$. In 1995, LEP concluded its $Z$ operations with an energy scan. Only the on-peak data from this scan, corresponding to an integrated luminosity of 15~pb$^{-1}$, are used here. The total integrated luminosity is 61~pb$^{-1}$. The DELPHI Open Data release~\cite{DELPHI:2024opendata} and data re-use policy~\cite{DELPHI:2024policy} make these datasets publicly available.

The primary MC samples used for the corrections are generated with \textsc{PYTHIA}~8.3~\cite{Bierlich:2022pfr} using the Monash 2013 tune~\cite{Skands:2014pea} and processed through the full DELPHI detector simulation, DELSIM~\cite{DELSIM}. The DELPHI detector and offline software differ slightly between the two years, denoted by the tags $94\_\text{c}$ and $95\_\text{d}$ in DELSIM. Separate samples of 5 million hadronic $Z \to q\bar{q}$ events per year are therefore produced for the two configurations, and the analysis is performed independently on the two datasets. Leptonic samples of $Z \to \tau^+\tau^-$, $Z \to \mu^+\mu^-$, and $Z \to e^+e^-$, of one million events each per year, are generated with the same \textsc{PYTHIA}~8.3 configuration and identical DELSIM detector simulation. These samples provide the signal and background estimates of the $\tau^+\tau^-$ control region (Section~\ref{sec:data_driven}). Two additional samples from the DELPHI legacy campaign~\cite{DELPHI:OpenData:kk2f_pythia_94, DELPHI:OpenData:kk2f_pythia_95}, which use \textsc{KK}2f~\cite{Jadach:1999vf} for the hard $e^+e^- \to q\bar{q}$ process and \textsc{PYTHIA}~6.1/\textsc{JETSET}~7.4~\cite{Sjostrand:2000wi} for parton showering and hadronization, are used to evaluate systematic uncertainties.

The baseline track selections follow those established in Refs.~\cite{DELPHI:2003yqh, DELPHI:2004wvq} and utilized in Ref.~\cite{Zhang:2025delphiEEC}. Charged particles are required to be contained within the polar-angle acceptance $20^\circ \le \theta \le 160^\circ$, to have a measured track length greater than 30~cm, and to satisfy a relative momentum uncertainty $\Delta p/p \le 1.0$. Badly reconstructed tracks concentrate at low polar angles, at large impact parameters, and at anomalously high momenta. Two further requirements are therefore tightened relative to the baseline to reduce charge misreconstruction. First, the track impact parameters are restricted to $|d_0| \le 0.6$~cm and $|z_0| \le 1.0$~cm, which removes displaced and secondary tracks whose charge is most prone to misreconstruction. Second, the track transverse momentum is raised to $p_{\rm T} > 2$~GeV. This cut suppresses soft tracks, whose charge is the most affected by misreconstruction, in particular from multiple Coulomb scattering at low momentum and low polar angle. It also improves the statistical sensitivity to the charge-asymmetry signal, since the charge-symmetric tracks from soft fragmentation add to the fluctuations of the correlator without adding to its mean, although the precision of the present measurement is set by the control sample rather than by the hadronic sample. Anomalously high-momentum tracks, in contrast, are automatically suppressed by the steeply falling momentum spectrum. Their residual contribution is assessed as a systematic uncertainty (Section~\ref{sec:systematicsSource}). Together, the tightened requirements also bring the signal-region track kinematics closer to those of the $\tau^+\tau^-$ control region used for the data-driven charge-misreconstruction correction (Section~\ref{sec:data_driven}). Hadronic events are selected by requiring at least 7 good charged tracks, total reconstructed energy $E_{\rm vis} \ge 0.5\,E_{\rm cm}$, and a thrust-axis polar angle in the range $30^\circ \le \theta_{\rm thrust} \le 150^\circ$. These selections remove background from leptonic final states and two-photon collisions, and suppress the tail of radiative-return events in which hard initial-state radiation lowers the effective annihilation energy $\sqrt{s'}$. Approximately 1.5 million hadronic events are selected across both years. A complete summary of the particle and event selections is given in Table~\ref{tab:SelectionSummary}.

\begin{table}[ht]
\centering
\begin{tabularx}{0.75\textwidth}{l|l}
\hline\hline
\multicolumn{2}{l}{Charged particles}  \\
\hline
Acceptance              & $20^\circ\le\theta\le160^\circ$ \\
Transverse momentum     & $p_{\rm T} > 2~\text{GeV}$ \\
High quality tracks     & measured track length $\ge 30~\text{cm}$ \\
                        & $\Delta p/p \le1.0 $ \\
Impact parameter        & $|d_0|\le0.6$~cm, $|z_0|\le1.0$~cm \\
\hline\hline
\multicolumn{2}{l}{Event selection}  \\
\hline
Hadronic events         & $30^\circ \le \theta_{\rm thrust} \le 150^\circ$\\
                        & at least 7 good tracks \\
                        & $E_{\rm vis} \ge 0.5\,E_{\rm cm}$ \\
\hline\hline
\end{tabularx}
\caption{Summary of particle and event selections used in this analysis.}
\label{tab:SelectionSummary}
\end{table}

Generator-level studies with 10 million \textsc{PYTHIA}~8.3 events confirm that the $\sin(2\theta)$ modulation anticipated in Section~\ref{sec:observable} survives hadronization and is preserved by the $p_{\rm T}$ requirement, whereas the thrust-axis polar-angle requirement of the event selection suppresses the modulation in the bins nearest the acceptance edges. A correction for this effect is derived from simulation and applied to the final results (Section~\ref{sec:results}).


\section{Simulation-based detector correction}
\label{sec:mc_correction}
 
\begin{figure}[ht!]
    \centering
    \includegraphics[width=0.49\textwidth]{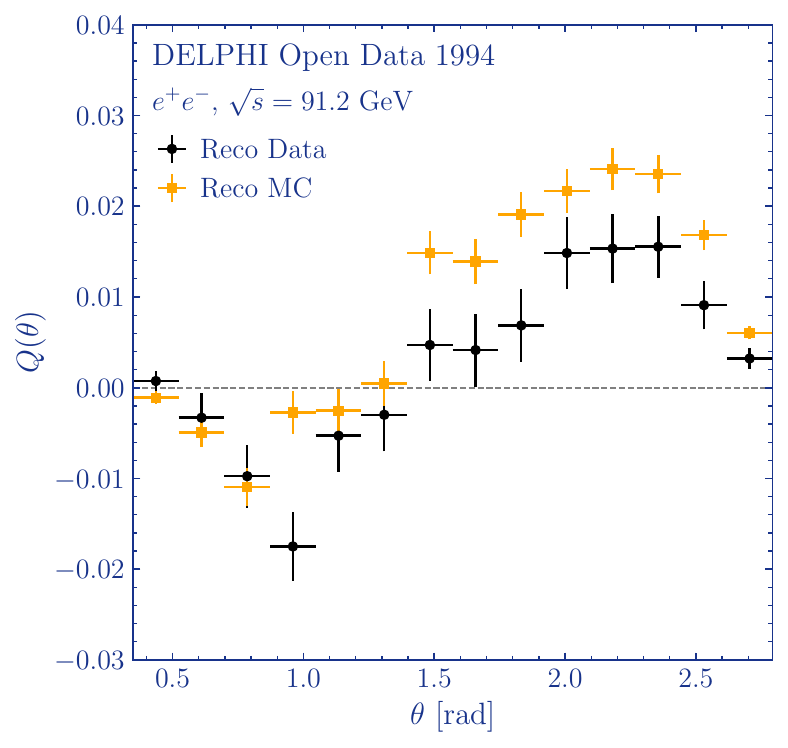}
    \includegraphics[width=0.49\textwidth]{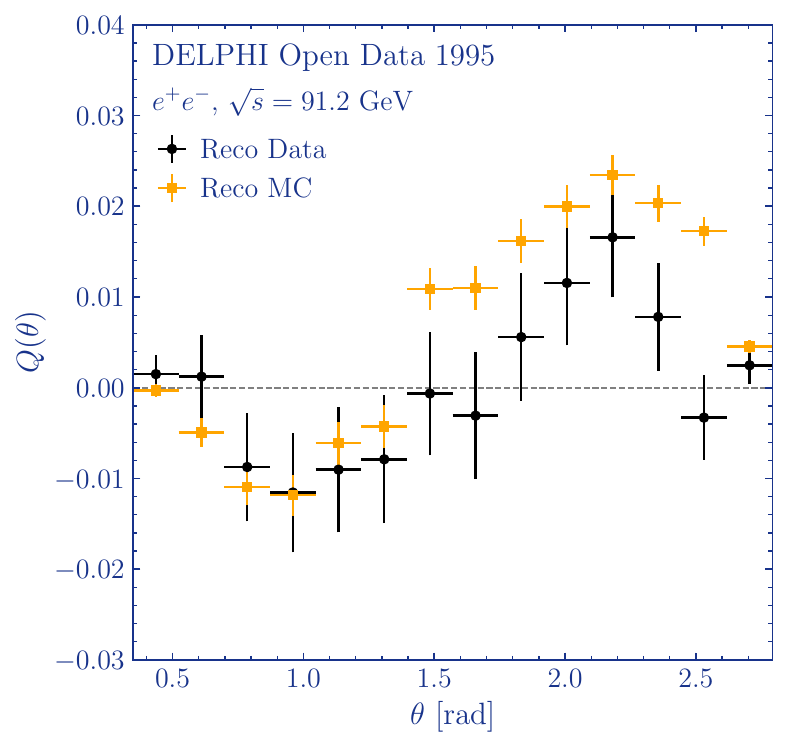}
    \caption{Comparison of the one-point charge correlator at detector level between data (black) and the reconstructed \textsc{PYTHIA}~8.3 simulation (yellow), before any correction for the 1994 (left) and 1995 (right) samples.}
    \label{fig:data_mc_fbcc}
\end{figure}

Comparisons between data and the reconstructed \textsc{PYTHIA}~8.3 simulation are performed for the key kinematic distributions used in this analysis and are documented in the accompanying analysis note. Figure~\ref{fig:data_mc_fbcc} shows the one-point charge correlator at detector level for the 1994 and 1995 datasets. A characteristic parity-odd modulation is clearly visible in both the data and the simulation already at this stage, before any correction, and the overall $\theta$ dependence is well reproduced. The reconstructed simulation, however, exhibits a small but clear excess of positive over negative tracks that is absent in the data and persists across the full $\theta$ range. The analysis therefore adopts a two-stage correction strategy. The charge-dependent reconstruction effects as modeled in the simulation, track finding inefficiency, fake tracks, and charge misidentification, are corrected with the simulation-based procedure described in this section. The residual data--simulation difference in the charge-reconstruction response, which no simulation-derived correction can capture, is then measured by the data-driven method of Section~\ref{sec:data_driven}.
 
The simulation-based correction is derived from the fully simulated samples of Section~\ref{sec:data_samples}. A correspondence between reconstructed and generator-level tracks is established with the Hungarian algorithm~\cite{hungarianMatching}, using the opening angle between the reconstructed and generator-level momenta as the cost metric and a maximum matching distance of 0.05~rad. Matched pairs are further classified by charge. A reconstructed track that is geometrically matched but carries the opposite charge to its generator-level partner is classified as charge misidentification. A reconstructed track with no geometric match is classified as fake, and a generator-level track with no reconstructed partner indicates track finding inefficiency. The resulting track finding efficiencies, fake rates, and misidentification rates are evaluated separately for the two charge signs. In the simulation, the track finding efficiency is approximately 85\% throughout the bulk of the acceptance and is slightly higher for positive than for negative tracks. The combined charge-misidentification and fake rate ranges from below 1\% to about 5\%, depending on $\theta$. The full decomposition of the reconstructed sample, by category and charge, is documented in the accompanying analysis note. Beyond providing the inputs to the correction, the same matching is used below to diagnose the origin of the charge imbalance seen in Figure~\ref{fig:data_mc_fbcc}.
 
The correction is applied to the data in two steps. First, the combined contribution of charge-misidentified and fake tracks, estimated from simulation, is subtracted from the reconstructed positive- and negative-track yields in each angular bin. Second, the resulting yields are corrected for same-charge track finding efficiency and converted to per-event track densities using the normalization of Eq.~\eqref{eq:charged_densities}. The simulation-corrected charge correlator, denoted $\mathcal Q_{\rm eff}(\theta)$, is obtained from the difference of the corrected positive- and negative-track densities.
 
The matching further enables a track-level diagnosis of the charge imbalance of the reconstructed simulation, which is decomposed by the flavor of the primary quark pair and by the final-state hadron species. Using the 1994 sample as an example, the $\theta$-integrated net imbalance per event, $Q_{\rm net}$, in which the parity-odd physics asymmetries largely cancel, has the same positive sign and comparable magnitude, $\mathcal{O}(10^{-3}\text{--}10^{-2})$, for the $b\bar{b}$, $c\bar{c}$, and light-quark classes, showing no dependence on the primary flavor. Decomposed by species, $Q_{\rm net}$ is statistically compatible with zero at generator level and becomes significantly positive after reconstruction: for pions it goes from $(0.47 \pm 2.23)\times10^{-3}$ to $(7.53 \pm 1.87)\times10^{-3}$, for kaons from $(-0.28 \pm 0.91)\times10^{-3}$ to $(14.78 \pm 0.79)\times10^{-3}$, and for protons from $(0.87 \pm 0.62)\times10^{-3}$ to $(31.13 \pm 0.48)\times10^{-3}$. Both the sign and the ordering across species follow the particle--antiparticle differences in the inelastic interaction cross sections with the detector material, which are smallest for pions and largest for protons, and these differences are included in the detector simulation. The sizable shift for pions, for which the particle--antiparticle cross-section difference is by far the smallest, suggests that species-independent reconstruction effects contribute.
 
The dependence of the simulation-based correction on the event generator is assessed by repeating the procedure with the legacy \textsc{PYTHIA}~6.1/\textsc{JETSET}~7.4 sample in place of the nominal \textsc{PYTHIA}~8.3 sample, and the difference is propagated as a systematic uncertainty (Section~\ref{sec:systematicsSource}).
When the correction is applied to the data, a clear global charge imbalance remains, indicating a residual data--simulation difference in the charge-reconstruction asymmetry that the simulation-based correction does not capture. The estimation and correction of this residual is described in Sections~\ref{sec:cp_correction} and~\ref{sec:data_driven}.


\section{Data-driven calibration strategy}
\label{sec:cp_correction}
The simulation-corrected correlator, $\mathcal{Q}_{\rm eff}(\theta)$, differs from the true one by the residual data--simulation charge bias $b(\theta)$,
\begin{equation}\label{eq:bias_decomposition}
    \mathcal{Q}_{\rm eff}(\theta) =
    \mathcal{Q}_{\rm true}(\theta) + b(\theta).
\end{equation}
Because the correlator is the small difference of two large track yields, even a small charge-asymmetric residual, amplified by the total track density, can shift the measured distribution by an amount comparable to the signal itself. Determining $b(\theta)$ from data is therefore the central task of the calibration.

The general strategy exploits how the bias transforms under the reflection $\theta \to 180^\circ\!-\theta$. CP invariance of inclusive hadronic $Z$ decays requires the density of positive hadrons at $\theta$ to equal the density of negative hadrons at $180^\circ\!-\theta$, so the true correlator is exactly odd under this reflection, and the fiducial region is symmetric about $\theta = 90^\circ$. Splitting the bias into its even and odd components under the same reflection, $b(\theta) = b_{\rm even}(\theta) + b_{\rm odd}(\theta)$, the two components can be constrained in fundamentally different ways. The even component can be removed in situ, using the hadronic sample itself, exactly and without free parameters (see Eq.~\eqref{eq:folding_result}). The odd component carries the same angular structure as the parity-odd signal and is invisible to any self-calibration of the hadronic sample. Constraining it requires an additional input from outside the hadronic sample, either a calibration of the charge-dependent detector response or an independent measurement of the charge misreconstruction rate.

The LEP and SLC measurements of the inclusive hadronic charge asymmetry~\cite{ALEPH:1991fba, ALEPH:1996qlh, DELPHI:1991mqi, OPAL:1992jsm, L3:1991gfs, L3:1998jet, SLD:1996gjt} are based on the difference of the forward and backward hemisphere charges, $\langle Q_{\rm F} - Q_{\rm B}\rangle$. The forward--backward-symmetric part of the detector response cancels in the difference, while the sum $\langle Q_{\rm F} + Q_{\rm B}\rangle$ monitors its size. \footnote{These analyses were self-calibrating in a second respect as well. The charge separation between the quark and antiquark hemispheres, which converts the measured charge flow into the quark-level asymmetry, was extracted from the data through the correlation of the two hemisphere charges, $\langle Q_{\rm F}\,Q_{\rm B}\rangle$, with the hemisphere correlation taken from simulation~\cite{ALEPH:1996qlh, OPAL:2002ddm}.} The same strategy is followed here. The symmetric component of the detector response is removed by folding the simulation-corrected distribution $\mathcal{Q}_{\rm eff}(\theta)$ about $\theta = 90^\circ$. The folded distribution $\mathcal{Q}_{\rm fold}(\theta)$ is defined as
\begin{equation}\label{eq:folding_correction}
    \mathcal{Q}_{\rm fold}(\theta) \;\equiv\;
    \frac{1}{2}\left[\mathcal{Q}_{\rm eff}(\theta)
    - \mathcal{Q}_{\rm eff}(180^\circ\!-\theta)\right],
    \qquad 90^\circ < \theta \le 160^\circ .
\end{equation}
Unlike the hemisphere charges, whose hemispheres are defined by the event thrust or sphericity axis, the folding acts on the polar angle of each track with respect to the $e^-$ beam direction, and no event axis enters. The true correlator, being exactly odd, is unchanged by this projection, while the bias is reduced to its odd part,
\begin{equation}\label{eq:folding_result}
    \mathcal{Q}_{\rm fold}(\theta) =
    \mathcal{Q}_{\rm true}(\theta)
    + \frac{1}{2}\left[b(\theta) - b(180^\circ\!-\theta)\right].
\end{equation}
The parity-even component of the bias cancels identically, independently of its mechanism and differentially in $\theta$. The folding is an exact, parameter-free projection. It introduces no additional systematic uncertainty and no control-sample statistical uncertainty.

\begin{figure}[ht!]
    \centering
    \includegraphics[width=0.55\textwidth]{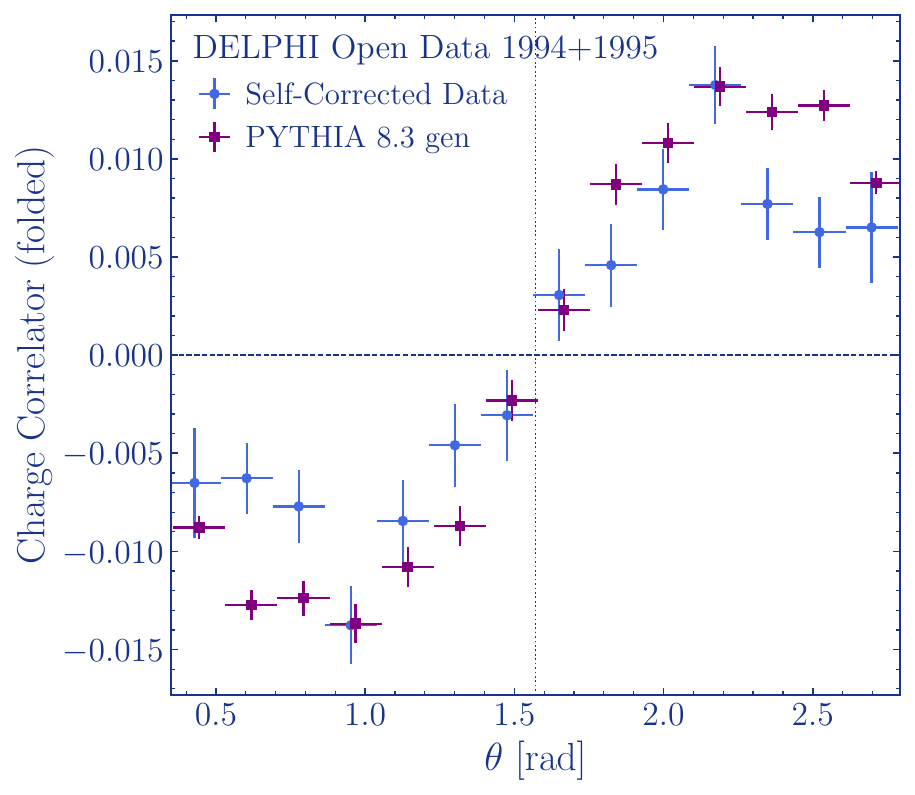}
    \caption{One-point charge correlator obtained with the self-calibration of Eq.~\eqref{eq:folding_correction}, combining the 1994 and 1995 DELPHI Open Data samples. The simulation-corrected distribution of Section~\ref{sec:mc_correction} is folded about $\theta = 90^\circ$, which removes the parity-even component of the residual charge bias. The folded \textsc{PYTHIA}~8.3 generator-level prediction is overlaid. The distributions are displayed over the full angular range by antisymmetric mirroring. Bins mirrored about $90^\circ$ are fully anti-correlated by construction. Error bars are statistical. The folded distribution serves as a cross-check of the nominal result, which uses the data-driven correction of Section~\ref{sec:data_driven}.} 
    \label{fig:folded_crosscheck}
\end{figure}

Figure~\ref{fig:folded_crosscheck} shows the folded correlator for the combined 1994 and 1995 data, compared with the correspondingly folded \textsc{PYTHIA}~8.3 generator-level prediction, with the shape change induced by the hadronic event selection corrected as described in Section~\ref{sec:data_samples}. The folding yields the distribution on the half-range $90^\circ < \theta \le 160^\circ$. It is displayed over the full angular range by antisymmetric mirroring, with mirrored bins fully anti-correlated by construction. A charge asymmetry consistent with the expected $\sin(2\theta)$ modulation is observed. The projection thus removes the parity-even part of the residual bias in situ, including the global charge imbalance observed in Section~\ref{sec:mc_correction}.
 
The parity-odd component of the bias, by contrast, survives the projection unchanged. Such a component consists of opposite net charge migrations at mirror angles. A net positive-to-negative migration at $\theta = 30^\circ$, accompanied by a net negative-to-positive migration at $\theta = 150^\circ$, lowers the measured correlator in one hemisphere and raises it in the other, and cannot be separated from the signal in situ.

Previous measurements controlled these parity-odd charge biases with dedicated detector studies. The early DELPHI, OPAL, and L3 measurements~\cite{DELPHI:1991mqi, OPAL:1992jsm, L3:1998jet} relied on extensive verification with the detector simulation. ALEPH supplemented the detector simulation with external calibrations, deriving a sagitta calibration from the momentum balance of $Z\to\mu^+\mu^-$ events and a calibration of the detector material from photon conversions~\cite{ALEPH:1991fba, ALEPH:1996qlh}. A later OPAL measurement on $b$-enriched samples likewise measured the forward--backward asymmetry of the detector material with photon conversions, finding a significant asymmetry only near the edge of the acceptance~\cite{OPAL:2002ddm}. At SLD, the charge-dependent forward--backward sagitta bias was studied with dimuon and Bhabha events and constituted the largest systematic uncertainty of the measurement~\cite{SLD:1996gjt}.

All of these approaches rest on extensive detector-level calibration and verification, and their counterparts for the DELPHI Open Data are natural next steps toward a precision measurement. As a first step in this direction, the forward--backward asymmetry of the detector material, probed with the rate of reconstructed photon conversions, and the track momentum scale and resolution, probed with the position and width of the $K^0_S \to \pi^+\pi^-$ mass peak, are compared between data and simulation as functions of the polar angle~\cite{Zhang:2026esd}. The forward--backward asymmetry of the detector material is well reproduced by the simulation, with residual data--simulation differences at the percent level and mostly confined to the bin containing the TPC cathode plane. The momentum scale agrees to within $10^{-3}$, and the momentum resolution is forward--backward symmetric to about 2\% in both data and simulation. A per-charge calibration of the detector response, in particular of the effects of the detector material and of the track sagitta, along the lines of the previous measurements, is left to that future work.

In this analysis, the residual charge bias is instead bounded with a single data-driven measurement: a tag-and-probe determination of data--simulation differences in the charge misreconstruction rate in $e^+e^- \to \tau^+\tau^-$ events, described in the next section. Measured as a function of the probe kinematics, the residual retains both parity components, so a single procedure addresses the parity-even and parity-odd parts of the detector bias at once. The precision of the calibration is limited by the size of the control sample and sets the dominant systematic uncertainty of the measurement (Section~\ref{sec:systematics}).


\section{Charge reconstruction measurement in the control region}
\label{sec:data_driven}
Measuring the charge misreconstruction rate in data requires a sample of tracks whose true charge is known independently of the reconstruction. The process $e^+e^- \to \tau^+\tau^-$ provides such a sample at the single-track level. Since $\tau$ leptons decay predominantly into low-multiplicity final states, charge conservation and topology requirements determine the true charge of the selected tracks with high purity, enabling a data-driven determination of the charge-misreconstruction rate.

\begin{figure}[ht!]
    \centering
    \includegraphics[width=0.48\textwidth]{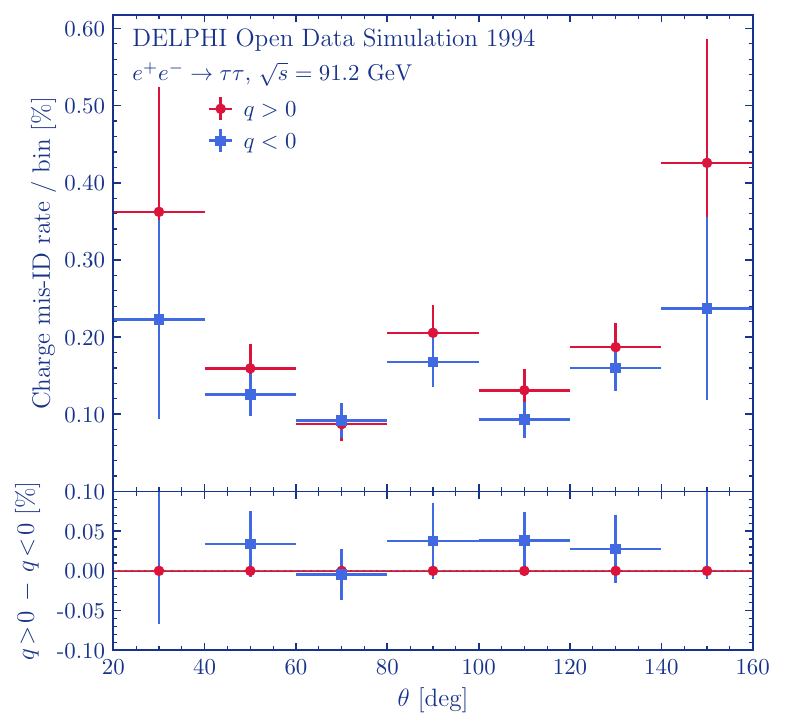}
    \includegraphics[width=0.48\textwidth]{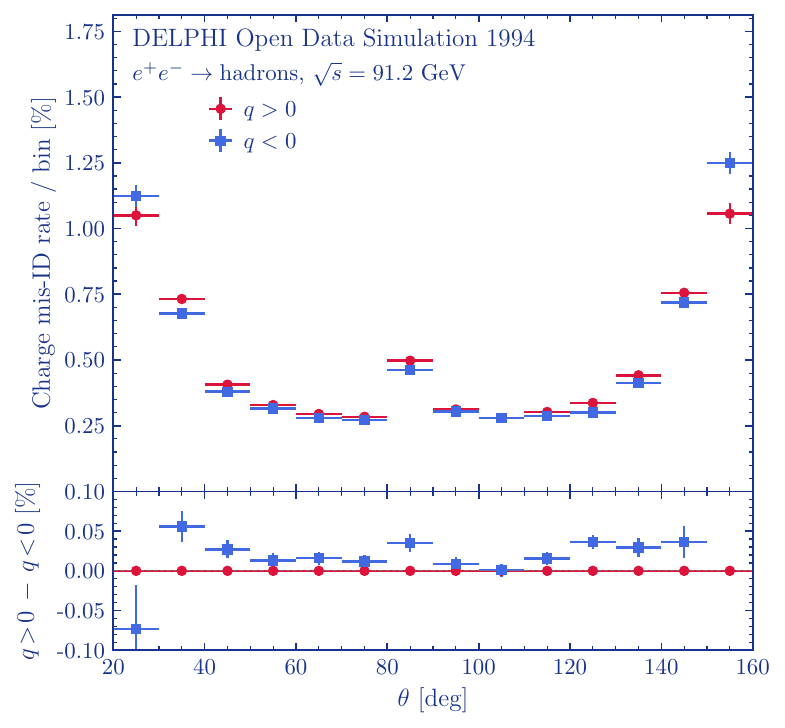}
    \caption{Per-track charge-misreconstruction rate obtained from truth matching in the \textsc{PYTHIA} 8.3 simulation, defined as the fraction of reconstructed tracks of a given charge whose matched generator-level partner carries the opposite charge, shown separately for the two reconstructed charges ($q>0$ and $q<0$), in the $\tau\tau$ control region (left) and in inclusive hadronic events (right). In both environments, the rate is larger for $q>0$, corresponding to a net negative-to-positive charge migration. The absolute rates are larger in the hadronic sample, while the difference between the rates for the two charges is consistent between the samples within the statistical precision.}
    \label{fig:tau_misid_rate_mc}
\end{figure}

The event environment in $\tau^+\tau^-$ decays differs from that of hadronic $Z$ decays. The track multiplicity is much lower, the tracks are more isolated, and the kinematic distributions differ. The transferability of the correction is examined with generator-level studies and truth matching. Figure~\ref{fig:tau_misid_rate_mc} compares the per-track charge-misreconstruction rates, evaluated separately for positive and negative tracks, in the $\tau^+\tau^-$ and inclusive hadronic samples for 1994. The absolute rates are larger in the hadronic sample, as expected from the higher track density. The difference between the negative- and positive-track rates, which drives the correction, is instead consistent between the two samples in magnitude and $\theta$ dependence within the statistical precision. The same behavior is observed in the 1995 sample. This matches the expectation that environment effects, such as track-finding confusion in dense events, are largely charge-symmetric, while the charge-asymmetric component arises from track-level interactions with the detector material and from residual distortions of the track geometry. This picture is also independently supported by the decomposition of the simulated charge imbalance by primary quark flavor and by final-state hadron species (Section~\ref{sec:mc_correction}). Importantly, the correction does not transfer the absolute misreconstruction rate from $\tau^+\tau^-$ to hadronic events. It measures the data--MC difference of the rate, which is the component arising from mismodeling of the track-level detector response and is common to the two event environments. Residual differences in the absolute rates between the environments do not enter the correction directly, since they are present in both data and simulation and cancel in the data--MC difference. The environment-sensitive component is moreover largely charge-symmetric and can therefore contribute only through the dilution term, whose impact is below the percent level (Section~\ref{sec:systematics}). 

\subsection{$Z \to \tau^+\tau^-$ control region}
\label{sec:tau_control}

The $\tau$ control sample is selected by requiring an event topology consistent with $\tau$-pair production. Charged tracks must satisfy the same baseline quality criteria as the main measurement (Table~\ref{tab:SelectionSummary}). The event must contain between 2 and 4 such tracks, with a visible energy fraction $0.10 < E_{\rm vis}/E_{\rm cm} < 0.80$: the low multiplicity rejects hadronic $Z$ decays, the lower energy bound suppresses two-photon events, and the upper bound suppresses high-visible-energy dilepton backgrounds, in particular Bhabha scattering and $Z \to \mu^+\mu^-$. The event thrust is required to exceed 0.985, and the acollinearity between the two thrust-hemisphere momenta must be less than $20^\circ$ to suppress QCD processes. The total visible mass is required to satisfy $15 < M_{\rm vis} < 55$~GeV. The upper bound provides a further handle on Bhabha and dimuon events, while the lower bound removes boosted two-photon events that survive the $E_{\rm vis}$ requirement but retain a small invariant mass. 

A further requirement based on particle identification applies a lepton veto, which suppresses the residual $Z \to e^+e^-$ and $Z \to \mu^+\mu^-$ contamination. The veto also suppresses $\tau$ decays to electrons and muons, thereby enriching the control sample in charged hadrons and bringing the track composition of the control region closer to that of the hadronic sample. All the selections are summarized in Table~\ref{tab:TauSelection}. The $\tau$-pair purity ($N_{\tau\tau}/N_{\rm sel}$) exceeds 96\% across the selected mass range, and the total leptonic purity ($1 - N_{q\bar{q}}/N_{\rm sel}$) exceeds 99\%, as estimated from simulation.

\begin{table}[ht]
\centering
\begin{tabularx}{0.75\textwidth}{l|l}
\hline\hline
\multicolumn{2}{l}{$\tau$-pair event selection}  \\
\hline
Track quality              & same as Table~\ref{tab:SelectionSummary} \\
Charged track multiplicity & $2 \le N_{\rm ch} \le 4$ \\
Visible energy fraction    & $0.10 < E_{\rm vis}/E_{\rm cm} < 0.80$ \\
Thrust                     & $T > 0.985$ \\
Acollinearity              & $< 20^\circ$ \\
Lepton veto                & veto events with identified $e$ or $\mu$\\
Visible mass               & $15 < M_{\rm vis} < 55$~GeV \\
\hline\hline
\end{tabularx}
\caption{Selection criteria for the $\tau$ control sample used in the charge-misreconstruction rate measurement.}
\label{tab:TauSelection}
\end{table}

\subsection{Tag-and-probe measurement of the misreconstruction rate}
\label{sec:tnp}

Within the selected $Z\to\tau\tau$ sample, the charge-misreconstruction rate is estimated using a tag-and-probe (TNP) method. The event is required to contain a pair of opposite-hemisphere tracks with one track in the central region $40^\circ \le \theta \le 140^\circ$, where the TPC delivers its full complement of 16 space points and the momentum and charge reconstruction are correspondingly most reliable. Given the acollinearity requirement, the other then lies in $20^\circ \le \theta \le 160^\circ$, matching the baseline track selection. About 23 thousand pairs are selected for the TNP. The $1{+}1$ prong topology selection purity, estimated from simulation, is above 98\%. Of the two tracks, the one with the smaller relative momentum uncertainty $\Delta p/p$ (the higher-quality momentum measurement) is chosen as the tag, and the other serves as the probe. Estimated using simulation, the charge-misreconstruction rate of the tag is below 0.08\% and that of the probe is around 0.2\% in the phase space of this analysis. Because the $\tau^+\tau^-$ pair is produced with zero net charge, the tag provides a high-purity reference for the expected charge of the probe: a negatively charged tag implies a positively charged probe, and vice versa.

The TNP charge-misreconstruction rate $f^{\pm}(\theta)$ is defined as the fraction of probe tracks whose reconstructed charge has the same sign as the tag, i.e.\ the same-sign fraction,
\begin{equation}\label{eq:misid}
    f^{\pm}(\theta) = \frac{N^{\rm SS}_{\pm}(\theta)}{N^{\rm tag}_{\pm}(\theta)},
\end{equation}
where $N^{\rm SS}_{\pm}(\theta)$ is the number of same-sign pairs and $N^{\rm tag}_{\pm}(\theta)$ the total number of tag entries, both as a function of the probe polar angle $\theta$ and binned by the true probe charge sign. Because the $\tau^+\tau^-$ pair carries zero net charge, the probe's true charge is fixed, opposite to that of the tag, by charge conservation alone, independently of the simulation. The same-sign fraction of Eq.~\eqref{eq:misid} therefore provides a direct, model-independent measurement of the probe charge-misreconstruction rate in both data and simulation, up to small residual tag misreconstruction and multi-prong topology leakage effects that are taken into account as systematic uncertainties.

The charge-misreconstruction rate is measured in four exclusive probe-$p_{\rm T}$ bins ($2$--$4$, $4$--$6$, $6$--$8$, and $>8$~GeV). Without loss of generality, we can reparametrize the data--MC residuals into an asymmetric combination $\delta_{\rm asym}(\theta)$ of $f^+$ and $f^-$, as well as a symmetric combination $\delta_{\rm sum}(\theta)$:
\begin{equation}
\delta_{\rm asym}(\theta) = (f^- - f^+)^{\rm data} - (f^- - f^+)^{\rm MC},
\qquad
\delta_{\rm sum}(\theta) = (f^+ + f^-)^{\rm data} - (f^+ + f^-)^{\rm MC}.
\end{equation}
In this way, the two combinations separate the two distinct effects of charge misreconstruction on the measurement. As shown in the next subsection, $\delta_{\rm asym}$ enters the corrected charge correlator multiplied by the total charged-particle density $N_{\rm tot}(\theta)$ defined in Eq.~\eqref{eq:ntot}, producing an additive bias, whereas $\delta_{\rm sum}$ multiplies the much smaller charge density $\mathcal Q(\theta)$ itself, producing only a dilution. Because $N_{\rm tot}(\theta)$ is much larger than $|\mathcal Q(\theta)|$, a residual of a given size has a far larger effect through the asymmetric term, which therefore drives the data-driven correction.

\begin{figure}[ht!]
    \centering
    \includegraphics[width=0.98\textwidth]{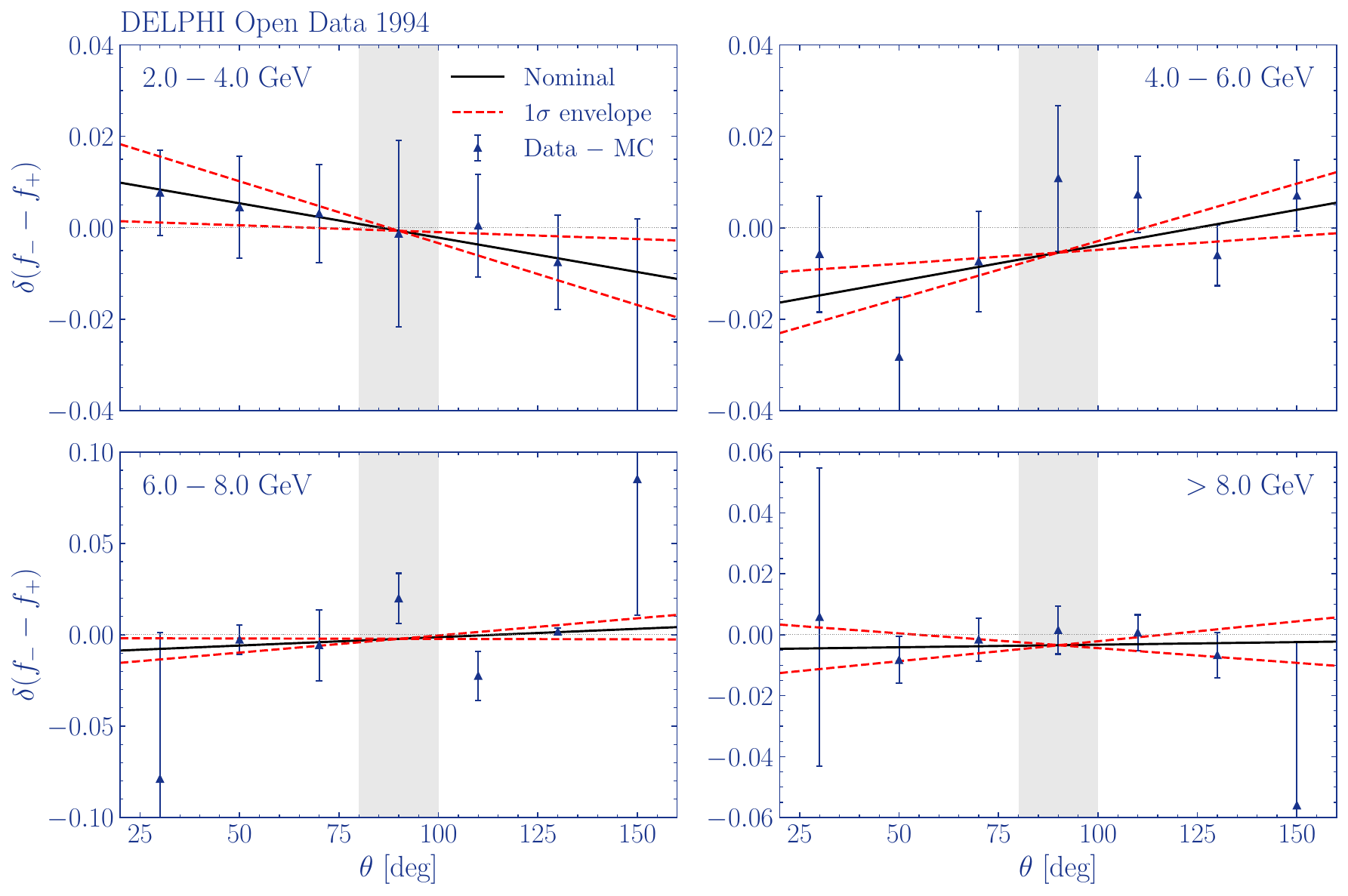}
    \caption{Total data--MC residual of the charge-misreconstruction asymmetry, $\delta_{\rm asym}(\theta) = \delta(f^- - f^+)$, as a function of the probe polar angle $\theta$ in the four exclusive probe-$p_{\rm T}$ bins entering the correction ($2$--$4$, $4$--$6$, $6$--$8$, and $>8$~GeV). Data points (blue triangles) show the residual with combined data and MC statistical uncertainties. The solid black line is the nominal first-order polynomial parametrization and the red dashed lines its $\pm 1\sigma$ envelope, obtained from the full covariance of the linear-fit parameters and propagated as the scale-factor uncertainty (Section~\ref{sec:systematics}). The shaded band between $80^\circ$ and $100^\circ$ marks the bins containing the DELPHI TPC cathode plane, which are excluded from the fit.}
    \label{fig:scale_factor}
\end{figure}

The data--MC residuals are extracted as a function of $\theta$ and parametrized by a first-order polynomial $\chi^2$ fit in the region $20^\circ \le \theta \le 160^\circ$, whose slope captures the parity-odd component of the charge-misreconstruction residual that mimics the signal, excluding the band $80^\circ \le \theta \le 100^\circ$ where a known TPC cathode-plane crack distorts the local charge response. The residuals and their parametrizations are shown in Figure~\ref{fig:scale_factor} in the four exclusive probe-$p_{\rm T}$ bins of the 1994 dataset. For most of the bins, the residuals are consistent with zero. The extracted residuals $\delta^i_{\rm asym}(\theta)$ are below 2\%. The fitted slopes are also statistically consistent with zero, within one standard deviation for the two higher $p_{\rm T}$ bins and within two standard deviations for the two lower bins, and they are mutually compatible, consistent with a purely parity-even residual. The residuals are also compatible between the two data-taking years. The full $2 \times 2$ covariance of the fit parameters is nevertheless retained in each probe-$p_{\rm T}$ bin and propagated downstream as scale-factor uncertainty, which is a conservative approach compared to assuming a purely parity-even residual and removing it only with the folding described in Section~\ref{sec:cp_correction}.

\subsection{Application of the correction}
\label{sec:misid_correction}

The dilution and the bias can be written explicitly in terms of the per-track sign-flip probabilities $f^{+}(\theta)$ and $f^{-}(\theta)$, the probabilities that a track of true positive or negative charge is reconstructed with the opposite sign, whose symmetric and asymmetric combinations are $\Sigma = f^+ + f^-$ and $\Delta = f^- - f^+$. In the following equations, the angular-bin index is suppressed, and $\mathcal Q$ and $N_{\rm tot}$ denote the binned densities defined in Eqs.~\eqref{eq:qNN} and~\eqref{eq:ntot}. The reconstructed and true correlators are then related by
\begin{equation}\label{eq:dilution}
    \mathcal{Q}_{\rm rec} = \mathcal{Q}_{\rm true}\,(1-\Sigma) + N_{\rm tot}\,\Delta,
\end{equation}
where the first term is the multiplicative dilution and the second the additive bias proportional to the total charged-particle density. The simulation-based correction of Section~\ref{sec:mc_correction} removes the charge migration predicted by the simulation. To first order in the residuals $\delta_{\rm sum} = \Sigma_{\rm data}-\Sigma_{\rm MC}$ and $\delta_{\rm asym} = \Delta_{\rm data}-\Delta_{\rm MC}$, the simulation-corrected correlator $\mathcal{Q}_{\rm eff}$ then obeys Eq.~\eqref{eq:dilution} with $\Sigma$ and $\Delta$ replaced by these residuals. Inverting that relation gives the data-driven correction,
\begin{equation}\label{eq:full_correction}
    \mathcal{Q}_{\rm corr}(\theta) =
    \frac{\mathcal{Q}_{\rm eff}(\theta) - \delta_{\rm asym}(\theta)\,N_{\rm tot}(\theta)}
         {1 - \delta_{\rm sum}(\theta)},
\end{equation}
where $\mathcal{Q}_{\rm eff}$ is the simulation-corrected charge correlator described in Section~\ref{sec:mc_correction}. The two residuals differ both in size and in the channel through
which they act. The symmetric residual $\delta_{\rm sum}$ is the larger of the two, being a sum rather than a difference of two misreconstruction rates. It is at the percent level and rescales $\mathcal{Q}_{\rm eff}$ by a relative amount of the same order. The asymmetric residual $\delta_{\rm asym}$ is smaller, but it is amplified by $N_{\rm tot}(\theta)$ as described in Section~\ref{sec:tnp}. It produces a shift comparable to $\mathcal{Q}$ itself and therefore carries the data-driven correction, dominating its uncertainty. The residuals $\delta_{\rm asym}(\theta)$ and $\delta_{\rm sum}(\theta)$ are obtained from the per-$p_{\rm T}$-bin $\tau$ measurements and reweighted to the hadronic-event $p_{\rm T}$ spectrum at each polar angle,
\begin{equation}\label{eq:pt_reweight}
    \delta_X(\theta) = \sum_i w^{\rm had}_i(\theta)\,\delta_X^i(\theta),
    \qquad
    w^{\rm had}_i(\theta) = \frac{N_{\rm tot}(p_{\rm T}\!\in\! i,\,\theta)}{N_{\rm tot}(\theta)},
    \qquad X \in \{\mathrm{asym},\,\mathrm{sum}\},
\end{equation}
where the index $i$ runs over the exclusive probe-$p_{\rm T}$ bins of Section~\ref{sec:tnp}, $\delta_X^i(\theta)$ is the residual measured in bin $i$, and $N_{\rm tot}(p_{\rm T}\!\in\! i,\,\theta)$ is the number of tracks in hadronic events with transverse momentum in bin $i$ at polar angle $\theta$, normalized to the total $N_{\rm tot}(\theta)$ at that angle. The weights satisfy $\sum_i w^{\rm had}_i(\theta) = 1$ by construction.

The procedure is validated end-to-end by a bias-injection closure test. A known charge-misreconstruction asymmetry, prescribed by the per-$p_{\rm T}$- and $\theta$-bin parametrized residuals $\delta = f^- - f^+$ measured on the 1994 sample, is injected at generator level by randomly flipping track charges in both the $\tau^+\tau^-$ control sample and the hadronic sample. The corrected charge correlator is then compared against the un-injected generator-level truth, in the high-statistics limit. In an idealized configuration restricted to truth-level $1{+}1$-prong $Z \to \tau^+\tau^-$ events, the per-$p_{\rm T}$-bin residuals are recovered by the linear parametrization within statistical precision, and the corrected hadronic correlator agrees with the un-injected truth, validating the algebraic inversion of Eq.~\eqref{eq:full_correction}. 
In a realistic configuration that includes $1{+}3$-prong decays in which two prongs are not reconstructed, a residual shape shift of $\sim 5\%$ remains after correction and is assigned as a systematic uncertainty on the data-driven procedure.

\begin{figure}[ht!]
    \centering
    \includegraphics[width=0.55\textwidth]{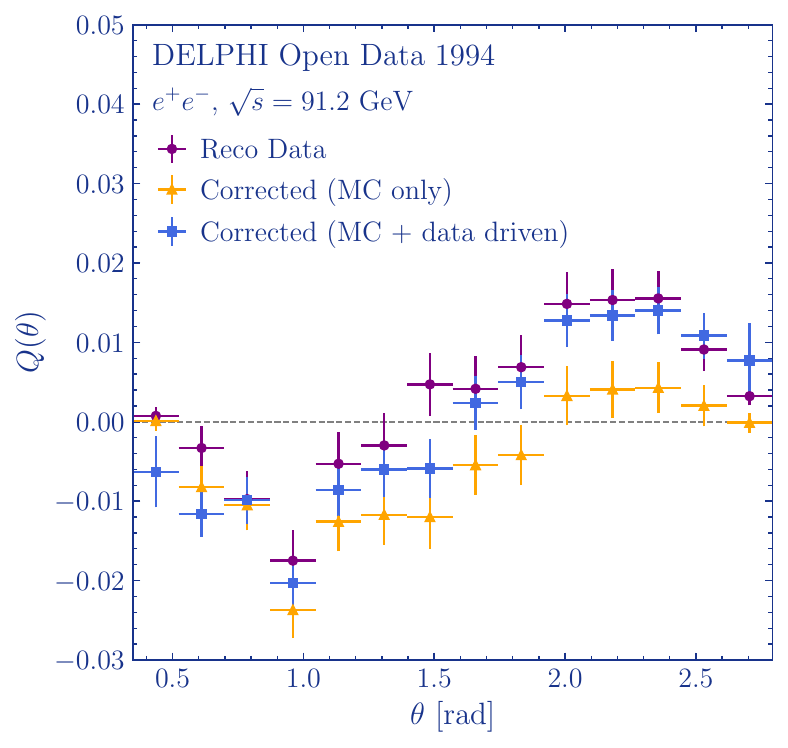}
    \caption{One-point charge correlator at three correction stages: the raw data (purple), MC-only corrected (orange), and fully corrected including the data-driven charge-misreconstruction correction (blue).}
    \label{fig:misid_corr}
\end{figure}

The full correction chain is shown in Figure~\ref{fig:misid_corr}, comparing the raw, simulation-corrected, and fully-corrected distributions for 1994. All three stages exhibit the parity-odd modulation. The simulation-based correction alone shifts the distribution downward across the full angular range, leaving a net negative charge. The data-driven step removes this offset, and the fully corrected distribution is approximately antisymmetric about $\theta = 90^\circ$, as expected by a parity-odd observable.


\section{Statistical and systematic uncertainties}
\label{sec:systematics}

The precision of this measurement is limited by sample size. Its dominant uncertainty is systematic, arising from the data-driven correction, and is itself driven primarily by the limited size of the control-region data sample. In this section, we describe details of the treatment of statistical and systematic uncertainties. 

\subsection{Statistical uncertainty}
\label{sec:stat}

Statistical uncertainties are evaluated from the event-by-event covariance of the per-bin net charge rather than from Poisson counting. This captures both the within-event track correlations, which inflate the variance relative to Poisson, and the anti-correlation between angular bins from charge conservation within the event, which reduces it. Because the efficiency correction is applied separately for the two charges, the covariance is accumulated per charge and propagated through the later MC-based and data-driven correction steps. The explicit expressions are given in the companion analysis note. The correction factors are held at their central values, and their statistical uncertainties enter as systematic uncertainties in Section~\ref{sec:systematicsSource}.

The 1994 and 1995 distributions and their statistical covariances are combined with weights given by the number of selected events in each year. The combined statistical covariance is added to the systematic covariance of Section~\ref{sec:syst_profile} to form the total covariance, and its contraction with the bin widths gives the statistical uncertainty on the integrated charge used in that section.

\subsection{Systematic uncertainty sources}
\label{sec:systematicsSource}

\paragraph{Anomalous high-momentum tracks:} A discrepancy exists between data and simulation in the high-$p_{\rm T}$ track spectrum. This systematic addresses the possibility that the data contain a source of high-$p_{\rm T}$ background tracks not modeled in the simulation. A $p_{\rm T}$-dependent fraction, ranging from 10\% to 40\% of charged tracks with $p_{\rm T} > 30$~GeV, is randomly removed from the data; the full difference to the nominal result is below 1\%.


\paragraph{MC model dependence:} The MC-based correction depends on the generator used to derive it. This is assessed by repeating the full correction with \textsc{PYTHIA}~6.1/\textsc{JETSET}~7.4 in place of the nominal \textsc{PYTHIA}~8.3 sample. The difference quantifies the dependence of the efficiency correction on the modeled track kinematics, hadron species composition, and event environment, and is propagated as the generator-model systematic.

\paragraph{Track finding efficiency:} The DELPHI collaboration tuned the simulated hit efficiencies to match data~\cite{Elsing:2000fv, Osterberg:1998xxx, DELPHI:1995dsm}, and a previous analysis~\cite{DELPHI:2000uri} estimated the residual efficiency uncertainty by randomly dropping reconstructed tracks at the 2\% level. We adopt the same random-drop technique, applied separately to positive and negative tracks, propagating the per-charge differences as independent shifts on $N^+$ and $N^-$. Because the two charges are varied independently, the resulting shift follows the total charged-particle density $N_{\rm tot}(\theta)$ rather than the much smaller charge correlator, and is therefore sizable and, to good approximation, symmetric under $\theta \to 180^\circ\!-\theta$. The constraint of Section~\ref{sec:syst_profile} bounds its angular integral but not its full angular shape, and a residual at the $10\%$ level remains. This is the second largest systematic source. 

\paragraph{Charge-misreconstruction bias scale factor:} The asymmetric residual $\delta_{\rm asym}(\theta)$ directly biases the charge correlator by introducing a spurious difference between the positive and negative track rates. Its statistical uncertainty is fully propagated via the covariance matrix of the linear function parameterization: in each of the four probe-$p_{\rm T}$ bins entering the correction ($2$--$4$, $4$--$6$, $6$--$8$, and $>8$~GeV), the $2 \times 2$ covariance is diagonalized into two orthogonal eigenmodes $\lambda_{+}$ and $\lambda_{-}$, yielding $2 \times 4 = 8$ independent shift vectors, treated as fully uncorrelated across both the eigenmode and probe-$p_{\rm T}$ indices. This is the dominant systematic source.

\paragraph{Charge-misreconstruction dilution scale factor:} The statistical uncertainty of the symmetric residual $\delta_{\rm sum}(\theta)$ is propagated in the same manner as the asymmetric residual, giving an additional 8 independent shift vectors. The impact on the corrected correlator is below 1\%, because $\delta_{\rm sum}$ enters as a dilution on the charge correlator distribution rather than being amplified by the much larger total number of tracks.

\paragraph{Tag-and-probe selection bias:} The TNP procedure carries residual selection biases from limited tag purity. A comparison between the TNP-extracted asymmetric residual and the corresponding generator-level truth in MC reveals a sub-percent discrepancy in the misreconstruction rate. Amplified by the total track yield, this translates into the third-largest systematic source.  

\paragraph{Tag-and-probe contamination modeling and bias:} A decomposition of the TNP residual by generator-level event topology shows that the dominant contribution comes from $1{+}3$-prong leakage into the nominal $1{+}1$ reconstructed sample. Although the leakage fraction is only $1.7\%$, the leaked events carry an intrinsically charge-asymmetric structure. A $10\%$ uncertainty on the $1{+}3$ topology contribution is propagated as a fully correlated shift ($\sim$1\%), and the residual non-closure of the bias-injection test (see Section~\ref{sec:misid_correction}) between the idealized and realistic configurations ($\sim$5\%) is assigned as the contamination bias.

\subsection{Application of the global charge-conservation constraint}
\label{sec:syst_profile}

We exploit global electric-charge conservation to constrain the systematic uncertainties.  The true charge correlator is odd under $\theta \to 180^\circ\!-\theta$, and the fiducial region is symmetric about $90^\circ$, so its angular integral over the measured range vanishes exactly. The deviation from zero observed in data therefore arises only from statistical fluctuations, of size $\sigma_{\rm stat}$, and from the parity-even component of the residual detector bias. Each of the systematic sources is instead fully correlated across $\theta$ and would shift the integrated charge well away from zero. The integrated charge measured in data therefore effectively bounds the normalization of each correlated systematic.

Let the corrected charge correlator be a vector $\mathbf{Q}$ of its values in the $N_\theta = 14$ bins of $\theta$. Each systematic source $i$ is propagated as a correlated $1\sigma$ shift $\mathbf{d}_i$, whose $\theta$ dependence is parametrized by a cubic function, applied uniformly to all sources to suppress bin-to-bin fluctuations while retaining the leading and next-to-leading terms in each parity sector. The parametrization acts primarily on the shifts derived from differences of finite simulated samples, as intended, and leaves the dominant charge-misreconstruction source unchanged to within $10\%$ in every angular bin. The systematic covariance is $V_{\rm syst} = \sum_{i} \mathbf{d}_i\,\mathbf{d}_i^{\!\top}$. The integrated charge is the scalar $I \equiv \mathbf{c}^{\!\top}\mathbf{Q}$, where $\mathbf{c}$ is the vector of bin widths that turns the $\theta$-binned correlator into the angular integral $\int \mathcal{Q}(\theta)\,\mathrm{d}\theta$. Charge conservation fixes the integral of the true correlator to zero, so the integrated charge observed in data, $I_{\rm obs}$, is accounted for by the integral of the correlated systematic offsets together with a statistical fluctuation of variance $\sigma_{\rm stat}^{2}$, the statistical uncertainty of the measured integral. Imposing this relation as a linear constraint in a least-squares fit of the systematic offsets yields the standard covariance update
\begin{equation}
\label{eq:schur}
V_{\rm syst} \;\longrightarrow\;
V_{\rm syst} \;-\;
\frac{V_{\rm syst}\,\mathbf{c}\,\mathbf{c}^{\!\top}\,V_{\rm syst}}
     {\mathbf{c}^{\!\top} V_{\rm syst}\,\mathbf{c}
      \;+\; \sigma_{\rm stat}^{2}} ,
\end{equation}
which reduces the covariance along the single direction $V_{\rm syst}\,\mathbf{c}$ resolved by the constraint and leaves all orthogonal directions unchanged. The limit $\sigma_{\rm stat} \to 0$ corresponds to an exact constraint, while $\sigma_{\rm stat} \to \infty$ removes it. The corresponding shift of the central values is proportional to $I_{\rm obs}$, which is consistent with zero within its statistical uncertainty, and is therefore not applied. Only the covariance update of Eq.~\eqref{eq:schur} is retained. The derivation is given in the analysis note~\cite{Zhang:2026esd}.

The charge-conservation constraint procedure acts primarily on the normalization component of the systematic shifts. A coherent shift with a non-zero angular integral is directly bounded by the measured integrated charge, whereas a shape variation aligned with the parity-odd signal, which integrates to zero over the symmetric fiducial range, is left largely unchanged. For the leading source, the charge-misreconstruction asymmetry residual, the constraint reduces the shift to approximately $67\%$ of its unconstrained value in the bins where the $\sin 2\theta$ modulation is largest. 

The systematic uncertainties are evaluated separately for the 1994 and 1995 datasets, following identical procedures, and each source is propagated as a single coherent shift, fully correlated across $\theta$ bins. Between the two years, the generator-model difference and the TNP sample-contamination modeling, which reflect physics modeling common to both datasets, are treated as fully correlated, while the remaining sources, driven by year-specific detector conditions and control-region statistics, are treated as uncorrelated.
The two years are combined with weights based on the number of selected events in each year. The shift vectors of the yearly correlated sources are combined with these weights before entering $V_{\rm syst}$, whereas the contributions of the yearly uncorrelated sources are added in quadrature with the same weights.

Table~\ref{tab:systematic_summary_9495} summarizes the systematic uncertainties of the combined result in the $\theta = 55^\circ$ and $\theta = 125^\circ$ bins, taken as representative of the forward and backward hemispheres in the angular region where the $\sin(2\theta)$ modulation and the size of the uncertainty are large. The quoted values are the square roots of the diagonal elements of the systematic covariance, expressed relative to the measured charge correlator in each bin. Each per-source entry is obtained by imposing the constraint on that source alone, whereas the total is obtained by imposing it on all sources simultaneously. The entries are therefore not additive. Their sum in quadrature does not reproduce the total. The systematic uncertainty is dominated by the statistical uncertainty of the charge-misreconstruction asymmetry residual correction factor, which contributes 25.9\% at $\theta = 55^\circ$ and 35.2\% at $\theta = 125^\circ$. Its size reflects the conservative choice of Section~\ref{sec:tnp} to propagate the full parametrization, including its parity-odd component, rather than assume a parity-even residual and remove it with the folding. The next-largest contribution is the charged-track efficiency, 10.3\% and 13.0\%. The corresponding shift follows the track density $N_{\rm tot}(\theta)$ and is, to good approximation, fully symmetric under $\theta \to 180^\circ\!-\theta$. The remaining contributions are the tag-and-probe selection bias (5.7\% and 6.6\%), the tag-and-probe contamination bias (4.6\% and 4.2\%), and the generator-model difference (0.7\% and 1.6\%). All other sources are below the $1\%$ level. The total systematic uncertainties, 27.6\% and 35.6\%, exceed the statistical uncertainties, 13.9\% and 18.2\%, by a factor of two in both bins. Because each source enters as a single coherent shift, its contribution to the covariance is fully correlated between angular bins, with the sign pattern set by the parity of the shift. Components even under $\theta \to 180^\circ\!-\theta$ correlate bins at mirror angles positively, while odd components anti-correlate them. Ultimately, the analysis is limited by the systematic uncertainty of the charge misreconstruction probability measurement, which is itself set by the size of the $\tau^+\tau^-$ control sample. 

\begin{table}[h]
\centering
\caption{Relative systematic uncertainties (\%) on the corrected one-point charge correlator for the combined 1994$+$1995 data, quoted in the $\theta = 55^\circ$ and $\theta = 125^\circ$ bins. The quoted values are the square roots of the diagonal elements of the systematic covariance, expressed relative to the measured charge correlator in each bin. Each per-source entry is obtained by imposing the constraint on that source alone, whereas the total is obtained by imposing it on all sources simultaneously. The entries are therefore not additive. Their sum in quadrature does not reproduce the total. Sources contributing below the percent level are not listed individually but are included in the total. The charge misreconstruction asymmetry residual is the combination of 16 uncorrelated components, eight per data-taking year.}
\label{tab:systematic_summary_9495}
\begin{tabular}{lcc}
\hline\hline
Source & $\theta = 55^\circ$ & $\theta = 125^\circ$ \\
\hline
Charged-track finding efficiency                     & 10.34 & 12.99 \\
Charge misreconstruction asymmetry residual  & 25.94 & 35.24 \\
Tag-and-probe selection bias                 & 5.67  & 6.58  \\
Tag-and-probe sample contamination bias      & 4.55  & 4.24  \\
MC generator model
                                             & 0.69  & 1.62  \\
\hline
Total systematic & 27.61 & 35.58 \\
Statistical      & 13.85 & 18.15 \\
\hline\hline
\end{tabular}
\end{table}

\section{Results}
\label{sec:results}

The one-point charge correlator is measured with DELPHI Open Data collected at $\sqrt{s} = 91.2$~GeV, for tracks with $p_{\rm T} > 2.0$~GeV in the polar-angle range $20^\circ \le \theta \le 160^\circ$. It is corrected for detector effects with the simulation-based correction described in Section~\ref{sec:mc_correction} and the data-driven charge misidentification correction of Section~\ref{sec:data_driven}, and for the shape change induced by the hadronic event selection described in Section~\ref{sec:data_samples}, derived from simulation. The 1994 and 1995 samples differ in detector configuration and are analyzed independently throughout, each with its own detector simulation and $\tau^+\tau^-$ control-region measurement.

\begin{figure}[ht!]
    \centering
    \includegraphics[width=0.55\textwidth]{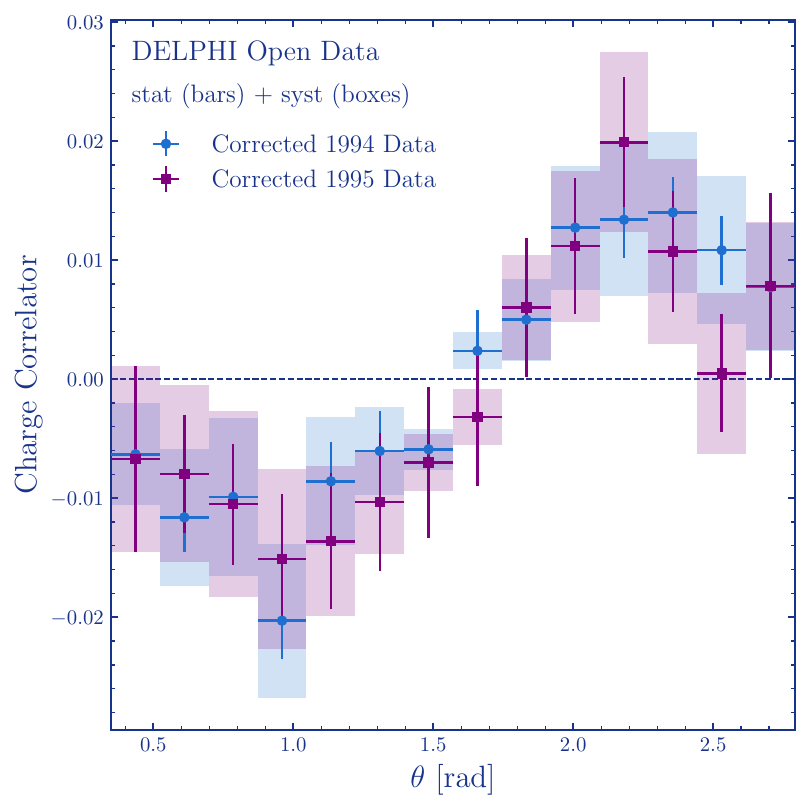}
    \caption{Fully corrected one-point charge correlator for tracks with $p_{\rm T} > 2.0$~GeV, comparing the DELPHI 1994 and 1995 samples. The two datasets are corrected independently, each with its own detector simulation and $\tau^+\tau^-$ control-region measurement. Data points are shown with statistical uncertainties (error bars) and systematic uncertainties (shaded boxes). The 1994 sample has an integrated luminosity of  46 pb$^{-1}$, whereas the 1995 sample has an integrated luminosity of 15 pb$^{-1}$, resulting in a lower statistical precision. }
    \label{fig:year_comparison}
\end{figure}

Figure~\ref{fig:year_comparison} compares the two corrected results. Both exhibit the characteristic $\sin(2\theta)$ modulation anticipated in Section~\ref{sec:observable}, with the same sign and comparable amplitude. The larger fluctuation of the 1995 sample reflects its smaller integrated luminosity, 15~pb$^{-1}$ against 46~pb$^{-1}$ in 1994. The two measurements are consistent within their uncorrelated uncertainties and are therefore combined. 
 
\begin{figure}[ht!]
    \centering
    \begin{overpic}[width=0.55\textwidth]{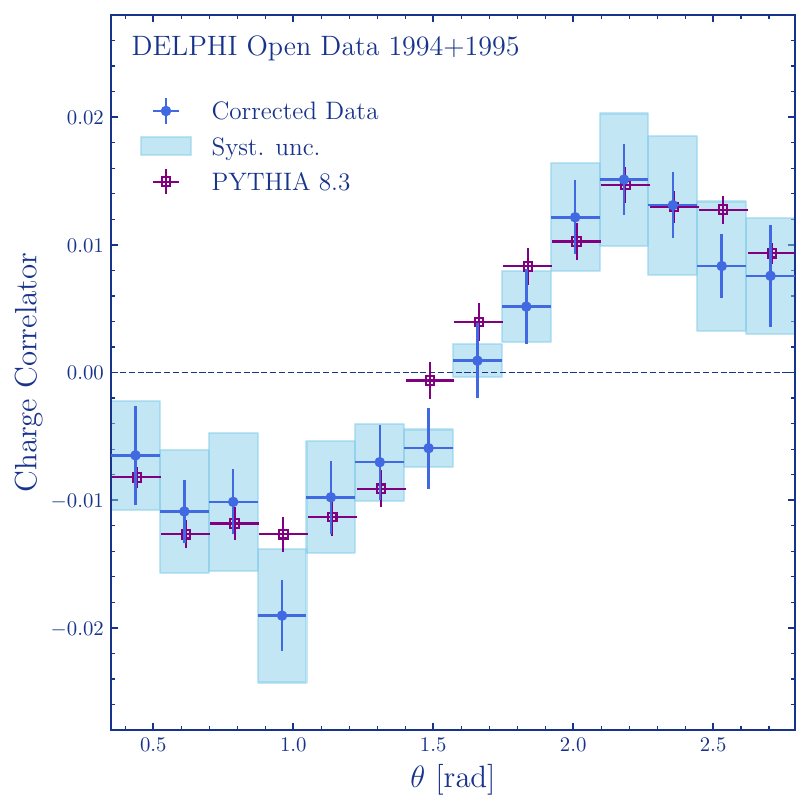}
        \put(64,15){\includegraphics[width=0.075\textwidth]{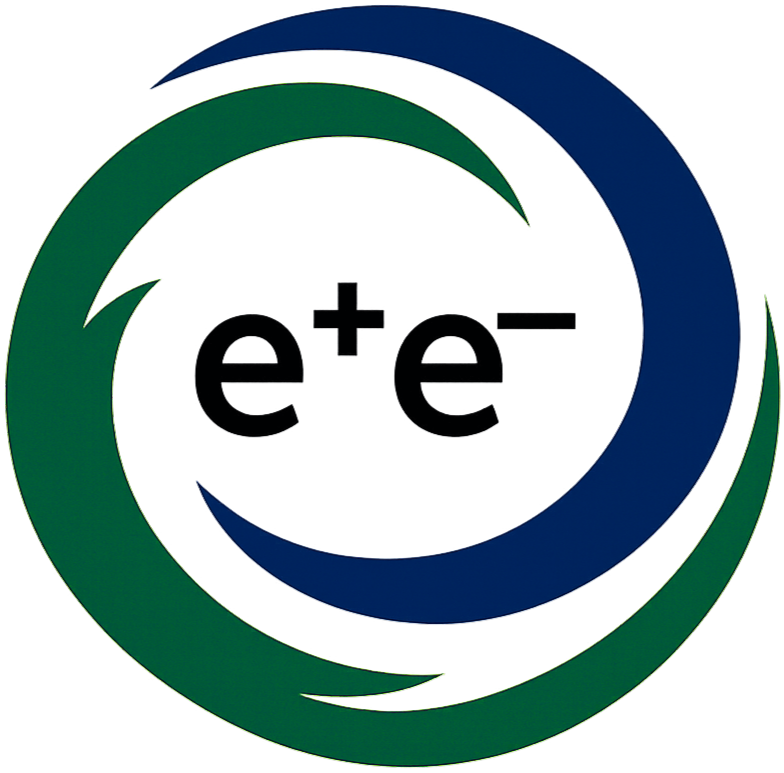}}
        \put(80,15){\includegraphics[width=0.075\textwidth,trim=129pt 52pt 204pt 45pt,clip]{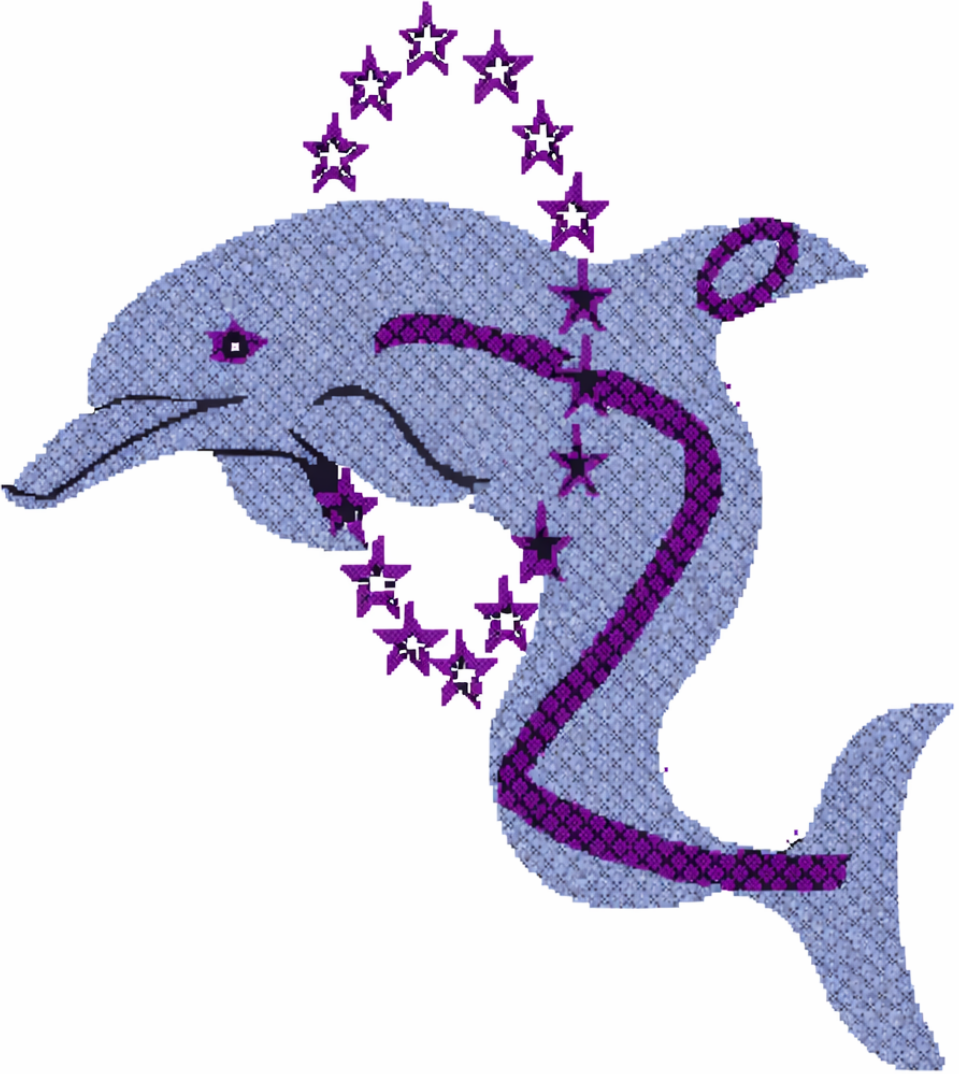}}
    \end{overpic}
    \caption{Fully corrected one-point charge correlator for tracks with $p_{\rm T} > 2.0$~GeV, combining the DELPHI 1994 and 1995 data, compared to the \textsc{PYTHIA}~8.3 generator-level prediction. Data points are shown with statistical uncertainties (error bars) and systematic uncertainties (shaded boxes). The data exhibit a clear $\sin(2\theta)$ modulation, consistent with the prediction within the evaluated uncertainties.}
    \label{fig:final_result_pythia}
\end{figure}

The combined result, weighted by the number of selected events in each year and with the between-year correlations described in Section~\ref{sec:syst_profile}, is shown in Figure~\ref{fig:final_result_pythia}. It is compared to the \textsc{PYTHIA}~8.3 generator-level prediction, evaluated with the same track momentum and polar-angle selections and including both initial-state radiation and $\gamma/Z$ interference. The one-point charge correlator is negative in the forward hemisphere and positive in the backward one, the sign expected from $\sum_q R_q Q_q A_{\rm FB}^q$, in which the down-type contributions dominate the sum, and it crosses zero at $\theta = 90^\circ$ as required by the symmetry. The measurement agrees with the prediction over the full angular range, within the evaluated uncertainties. 

This measurement establishes the experimental viability of the observable rather than providing a precision test of the Standard Model. Precision extraction of the hadronic asymmetries is not yet possible, but the measurement motivates the experimental and theoretical developments that such an extraction would require. On the experimental side, these are a precise per-charge calibration of the detector response in place of the tag-and-probe bound of Section~\ref{sec:data_driven}, and on the theoretical side, control of the non-perturbative correction that relates the hadronic charge flux to the quark-level asymmetry.

\section{Summary and outlook}
\label{sec:summary}

We report the first measurement of the one-point charge correlator in $e^+e^-$ collisions at the $Z$ pole, using $61~\mathrm{pb}^{-1}$ of archival DELPHI Open Data recorded at $\sqrt{s} = 91.2$~GeV in 1994 and 1995. The observable is the flux of electric charge into the detector as a function of the polar angle with respect to the $e^-$ beam direction. Unlike the jet-charge observables on which the previous hadronic asymmetry measurements were built, it involves no jets, no thrust hemispheres, and no event-by-event association of a reconstructed object with a primary-quark direction, and it resolves the parity-odd angular structure of hadronic $Z$ decays. The measurement is performed for tracks with $p_{\rm T} > 2$~GeV in $20^\circ \le \theta \le 160^\circ$.

The detector effects are corrected in two stages. First, a simulation-based correction is derived and applied separately to positive- and negative-charged tracks. Second, the residual data--simulation charge reconstruction differences are bounded by a tag-and-probe approach in the single-prong $e^+e^-\to Z\to\tau^+\tau^-$ sample, where charge conservation fixes the true charge of the probe track. The result, measured in bins of probe $p_{\rm T}$ and $\theta$, is transferred to the hadronic events. The parity-odd component of the residual, which is aligned with the angular structure of the signal, is statistically consistent with zero but is propagated in full. The data-driven charge misreconstruction measurement procedure therefore gives the dominant systematic uncertainty, whose size is set by the statistics of the control sample. The fully corrected one-point charge correlator exhibits the characteristic $\sin(2\theta)$ modulation and agrees with the \textsc{PYTHIA}~8.3 prediction.

This measurement establishes that the charge flux can be measured directly, and it identifies what a precision measurement requires. The precision is set by the size of the data sample together with the charge-dependent detector response, whose parity-odd component is the more difficult to control. Here it is bounded rather than assumed to vanish, by measuring the charge-misreconstruction rate in the control region. That rate is a binary per-track outcome, so the precision is set by the control-sample statistics. Direct measurements of the charge-sensitive tracking detector response, such as a sagitta calibration or a determination of the detector material asymmetry from photon conversions, are therefore the natural next step. Both follow the previous LEP and SLC standard without sample-size limitation of the present measurement, and both are within the reach of the DELPHI Open Data, whose release provides the simulation and reconstruction software alongside the data. 

While the hadronic forward--backward asymmetries were measured extensively at LEP and SLC with jet charge, the hadronic charge flux that the asymmetries encode had never been directly measured. Establishing that the asymmetry is accessible in charge flux, and identifying what limits its precision, opens a program of hadronic charge-flux measurements on archival $e^+e^-$ data. The same construction extends to fluxes of other hadronic currents, including heavy-flavor currents~\cite{Lee:2026bcq}, offering a direct route to the flavor-tagged asymmetries where the tensions are largest~\cite{Baak:2014ora, Haller:2018nnx, Fischer:2026bka}. Looking further ahead, the present dataset is a testing ground for the techniques that FCC-ee~\cite{FCC:2018byv} and CEPC~\cite{CEPCStudyGroup:2018ghi} will require, where Tera-$Z$ samples will remove the statistical limitation entirely and probe electroweak physics with unprecedented precision. The work is a first step in that direction.

\acknowledgments
This work would not have been possible without the decades of effort by the DELPHI collaboration in designing, building, and operating the detector, nor without the foresight of its data-preservation team. The authors are profoundly grateful for the DELPHI collaboration's monumental effort in making these pristine datasets and software infrastructure publicly available, and especially thank Dietrich Liko and Ulrich Schwickerath for their guidance on the DELPHI Open Data. J.Z. is deeply grateful to Sang Hyun Ko and Sitian Qian for their substantial work on the software stack for DELPHI Open Data analysis, on which this measurement partly relies. We thank Anthony Badea for reviewing the companion analysis note. We thank KITP Santa Barbara for its hospitality while this project was initiated. Y.C. and J.Z. are supported in part by the U.S. National Science Foundation under ID 2514008 and the Department of Energy grant DE-SC0026749.
Y.-J. Lee is supported by the Department of Energy, Office of Science, under Grant No. DE-SC0011088.

\bibliographystyle{JHEP}
\bibliography{paper-v4}

\providecommand{\href}[2]{#2}\begingroup\raggedright\begin{thebibliography}{10}

\bibitem{ALEPH:2005ab}
{\scshape ALEPH, DELPHI, L3, OPAL, SLD, LEP Electroweak Working Group, SLD
  Electroweak Group, SLD Heavy Flavour Group} collaboration, S.~Schael et~al.,
  \emph{{Precision electroweak measurements on the $Z$ resonance}},
  \href{https://doi.org/10.1016/j.physrep.2005.12.006}{\emph{Phys. Rept.}
  {\bfseries 427} (2006) 257--454},
  [\href{https://arxiv.org/abs/hep-ex/0509008}{{\ttfamily hep-ex/0509008}}].

\bibitem{Baak:2014ora}
{\scshape Gfitter Group} collaboration, M.~Baak, J.~C{\'u}th, J.~Haller,
  A.~Hoecker, R.~Kogler, K.~M{\"o}nig et~al., \emph{{The global electroweak fit
  at NNLO and prospects for the LHC and ILC}},
  \href{https://doi.org/10.1140/epjc/s10052-014-3046-5}{\emph{Eur. Phys. J. C}
  {\bfseries 74} (2014) 3046},
  [\href{https://arxiv.org/abs/1407.3792}{{\ttfamily 1407.3792}}].

\bibitem{Haller:2018nnx}
J.~Haller, A.~Hoecker, R.~Kogler, K.~M{\"o}nig, T.~Peiffer and J.~Stelzer,
  \emph{{Update of the global electroweak fit and constraints on
  two-Higgs-doublet models}},
  \href{https://doi.org/10.1140/epjc/s10052-018-6131-3}{\emph{Eur. Phys. J. C}
  {\bfseries 78} (2018) 675},
  [\href{https://arxiv.org/abs/1803.01853}{{\ttfamily 1803.01853}}].

\bibitem{Fischer:2026bka}
Y.~Fischer, J.~Haller, A.~Hoecker, R.~Kogler, F.~Labe, K.~M{\"o}nig et~al.,
  \emph{{The Higgs boson through the lens of electroweak precision data}},
  \href{https://arxiv.org/abs/2607.09861}{{\ttfamily 2607.09861}}.

\bibitem{Choudhury:2001hs}
D.~Choudhury, T.~M.~P. Tait and C.~E.~M. Wagner, \emph{{Beautiful mirrors and
  precision electroweak data}},
  \href{https://doi.org/10.1103/PhysRevD.65.053002}{\emph{Phys. Rev. D}
  {\bfseries 65} (2002) 053002},
  [\href{https://arxiv.org/abs/hep-ph/0109097}{{\ttfamily hep-ph/0109097}}].

\bibitem{Agashe:2006at}
K.~Agashe, R.~Contino, L.~Da~Rold and A.~Pomarol, \emph{{A Custodial symmetry
  for $Zb \bar b$}},
  \href{https://doi.org/10.1016/j.physletb.2006.08.005}{\emph{Phys. Lett. B}
  {\bfseries 641} (2006) 62--66},
  [\href{https://arxiv.org/abs/hep-ph/0605341}{{\ttfamily hep-ph/0605341}}].

\bibitem{OPAL:1997tsq}
{\scshape OPAL} collaboration, K.~Ackerstaff et~al., \emph{{Measurement of the
  branching fractions and forward - backward asymmetries of the Z0 into light
  quarks}}, \href{https://doi.org/10.1007/s002880050563}{\emph{Z. Phys. C}
  {\bfseries 76} (1997) 387--400},
  [\href{https://arxiv.org/abs/hep-ex/9707019}{{\ttfamily hep-ex/9707019}}].

\bibitem{OPAL:1992jsm}
{\scshape OPAL} collaboration, P.~D. Acton et~al., \emph{{A Measurement of the
  forward-backward charge asymmetry in hadronic decays of the $Z^0$}},
  \href{https://doi.org/10.1016/0370-2693(92)91546-L}{\emph{Phys. Lett. B}
  {\bfseries 294} (1992) 436--450}.

\bibitem{ALEPH:1991fba}
{\scshape ALEPH} collaboration, D.~Decamp et~al., \emph{{Measurement of charge
  asymmetry in hadronic Z decays}},
  \href{https://doi.org/10.1016/0370-2693(91)90844-G}{\emph{Phys. Lett. B}
  {\bfseries 259} (1991) 377--388}.

\bibitem{ALEPH:1996qlh}
{\scshape ALEPH} collaboration, D.~Buskulic et~al., \emph{{Determination of
  $\sin^2\theta_{\rm w}^{\rm eff}$ using jet charge measurements in hadronic
  $Z$ decays}}, \href{https://doi.org/10.1007/s002880050183}{\emph{Z. Phys. C}
  {\bfseries 71} (1996) 357--378}.

\bibitem{ALEPH:1998pmr}
{\scshape ALEPH} collaboration, R.~Barate et~al., \emph{{Determination of
  A-b(FB) using jet charge measurements in Z decays}},
  \href{https://doi.org/10.1016/S0370-2693(98)00345-1}{\emph{Phys. Lett. B}
  {\bfseries 426} (1998) 217--230}.

\bibitem{L3:1991gfs}
{\scshape L3} collaboration, B.~Adeva et~al., \emph{{Measurement of electroweak
  parameters from hadronic and leptonic decays of the $Z^0$}},
  \href{https://doi.org/10.1007/BF01475788}{\emph{Z. Phys. C} {\bfseries 51}
  (1991) 179--204}.

\bibitem{L3:1998jet}
{\scshape L3} collaboration, M.~Acciarri et~al., \emph{{Measurement of the
  effective weak mixing angle by jet-charge asymmetry in hadronic decays of the
  $Z$ boson}}, \href{https://doi.org/10.1016/S0370-2693(98)01174-5}{\emph{Phys.
  Lett. B} {\bfseries 439} (1998) 225--236}.

\bibitem{DELPHI:1991mqi}
{\scshape DELPHI} collaboration, P.~Abreu et~al., \emph{{A Measurement of
  $\sin^2\theta_{\rm W}$ from the charge asymmetry of hadronic events at the
  $Z^0$ peak}}, \href{https://doi.org/10.1016/0370-2693(92)90760-2}{\emph{Phys.
  Lett. B} {\bfseries 277} (1992) 371--382}.

\bibitem{OPAL:1993wua}
{\scshape OPAL} collaboration, R.~Akers et~al., \emph{{A Measurement of the
  forward - backward asymmetry of e+ e- ---{\ensuremath{>}} c anti-c and e+ e-
  ---{\ensuremath{>}} b anti-b at center-of-mass energies on and near the Z0
  peak using D*+- mesons}}, \href{https://doi.org/10.1007/BF01558389}{\emph{Z.
  Phys. C} {\bfseries 60} (1993) 601--612}.

\bibitem{ALEPH:2001mdb}
{\scshape ALEPH} collaboration, A.~Heister et~al., \emph{{Measurement of
  A**b(FB) using inclusive b hadron decays}},
  \href{https://doi.org/10.1007/s100520100812}{\emph{Eur. Phys. J. C}
  {\bfseries 22} (2001) 201--215},
  [\href{https://arxiv.org/abs/hep-ex/0107033}{{\ttfamily hep-ex/0107033}}].

\bibitem{L3:1992fsb}
{\scshape L3} collaboration, O.~Adriani et~al., \emph{{Measurement of the e+ e-
  --{\ensuremath{>}} b anti-b and e+ e- --{\ensuremath{>}} c anti-c forward
  backward asymmetries at the Z0 resonance}},
  \href{https://doi.org/10.1016/0370-2693(92)91203-L}{\emph{Phys. Lett. B}
  {\bfseries 292} (1992) 454--462}.

\bibitem{DELPHI:2004wvq}
{\scshape DELPHI} collaboration, J.~Abdallah et~al., \emph{{Determination of
  $A_{\rm FB}^{b}$ at the $Z$ pole using inclusive charge reconstruction and
  lifetime tagging}},
  \href{https://doi.org/10.1140/epjc/s2004-02104-0}{\emph{Eur. Phys. J. C}
  {\bfseries 40} (2005) 1--25},
  [\href{https://arxiv.org/abs/hep-ex/0412004}{{\ttfamily hep-ex/0412004}}].

\bibitem{OPAL:2002ddm}
{\scshape OPAL} collaboration, G.~Abbiendi et~al., \emph{{Measurement of the
  $b$ quark forward-backward asymmetry around the $Z^0$ peak using an inclusive
  tag}}, \href{https://doi.org/10.1016/S0370-2693(02)02594-7}{\emph{Phys. Lett.
  B} {\bfseries 546} (2002) 29--47},
  [\href{https://arxiv.org/abs/hep-ex/0209076}{{\ttfamily hep-ex/0209076}}].

\bibitem{L3:2000vgx}
{\scshape L3} collaboration, M.~Acciarri et~al., \emph{{Measurements of cross
  sections and forward-backward asymmetries at the $Z$ resonance and
  determination of electroweak parameters}},
  \href{https://doi.org/10.1007/s100520050001}{\emph{Eur. Phys. J. C}
  {\bfseries 16} (2000) 1--40},
  [\href{https://arxiv.org/abs/hep-ex/0002046}{{\ttfamily hep-ex/0002046}}].

\bibitem{SLD:2005gev}
{\scshape SLD} collaboration, K.~Abe et~al., \emph{{Direct measurements of
  $A_{b}$ and $A_{c}$ using vertex and kaon charge tags at SLD}},
  \href{https://doi.org/10.1103/PhysRevLett.94.091801}{\emph{Phys. Rev. Lett.}
  {\bfseries 94} (2005) 091801},
  [\href{https://arxiv.org/abs/hep-ex/0410042}{{\ttfamily hep-ex/0410042}}].

\bibitem{DELPHI:1994aml}
{\scshape DELPHI} collaboration, P.~Abreu et~al., \emph{{First measurement of
  the strange quark asymmetry at the Z0 peak}}, {\emph{Z. Phys. C} {\bfseries
  67} (1995) 1--14}.

\bibitem{SLD:2000jop}
{\scshape SLD} collaboration, K.~Abe et~al., \emph{{First direct measurement of
  the parity violating coupling of the Z0 to the s quark}},
  \href{https://doi.org/10.1103/PhysRevLett.85.5059}{\emph{Phys. Rev. Lett.}
  {\bfseries 85} (2000) 5059--5063},
  [\href{https://arxiv.org/abs/hep-ex/0006019}{{\ttfamily hep-ex/0006019}}].

\bibitem{DELPHI:1999mkl}
{\scshape DELPHI} collaboration, P.~Abreu et~al., \emph{{Measurement of the
  strange quark forward backward asymmetry around the Z0 peak}},
  \href{https://doi.org/10.1007/s100520000378}{\emph{Eur. Phys. J. C}
  {\bfseries 14} (2000) 613--631}.

\bibitem{ALEPH:2010aa}
{ALEPH, CDF, D0, DELPHI, L3, OPAL, and SLD Collaborations, the LEP Electroweak
  Working Group, the Tevatron Electroweak Working Group, and the SLD
  Electroweak and Heavy Flavour Groups}, \emph{{Precision Electroweak
  Measurements and Constraints on the Standard Model}},
  \href{https://arxiv.org/abs/1012.2367}{{\ttfamily 1012.2367}}.

\bibitem{Field:1977fa}
R.~D. Field and R.~P. Feynman, \emph{{A Parametrization of the Properties of
  Quark Jets}}, \href{https://doi.org/10.1016/0550-3213(78)90016-9}{\emph{Nucl.
  Phys. B} {\bfseries 136} (1978) 1--76}.

\bibitem{Catani:1999nf}
S.~Catani and M.~H. Seymour, \emph{{Corrections of O (alpha-S**2) to the
  forward backward asymmetry}},
  \href{https://doi.org/10.1088/1126-6708/1999/07/023}{\emph{JHEP} {\bfseries
  07} (1999) 023}, [\href{https://arxiv.org/abs/hep-ph/9905424}{{\ttfamily
  hep-ph/9905424}}].

\bibitem{Weinzierl:2006yt}
S.~Weinzierl, \emph{{The Forward-backward asymmetry at NNLO revisited}},
  \href{https://doi.org/10.1016/j.physletb.2006.11.076}{\emph{Phys. Lett. B}
  {\bfseries 644} (2007) 331--335},
  [\href{https://arxiv.org/abs/hep-ph/0609021}{{\ttfamily hep-ph/0609021}}].

\bibitem{dEnterria:2018jsx}
D.~d'Enterria and C.~Yan, \emph{{Forward-backward $b$-quark asymmetry at the Z
  pole: QCD uncertainties redux}},  in \emph{{53rd Rencontres de Moriond on QCD
  and High Energy Interactions}}, pp.~253--257, 2018,
  \href{https://arxiv.org/abs/1806.00141}{{\ttfamily 1806.00141}}.

\bibitem{Hofman:2008ar}
D.~M. Hofman and J.~Maldacena, \emph{{Conformal collider physics: Energy and
  charge correlations}},
  \href{https://doi.org/10.1088/1126-6708/2008/05/012}{\emph{JHEP} {\bfseries
  05} (2008) 012}, [\href{https://arxiv.org/abs/0803.1467}{{\ttfamily
  0803.1467}}].

\bibitem{Moult:2025nhu}
I.~Moult and H.~X. Zhu, \emph{{Energy Correlators: A Journey From Theory to
  Experiment}},  \href{https://arxiv.org/abs/2506.09119}{{\ttfamily
  2506.09119}}.

\bibitem{Zhang:2026emt}
J.~Zhang, K.~Lee, I.~Moult, Y.~Chen and Y.-J. Lee, \emph{{Observing Macroscopic
  Consequences of Electroweak Anomalies with Archival DELPHI Data}},
  \href{https://arxiv.org/abs/2608.14779}{{\ttfamily 2608.14779}}.

\bibitem{Electron-PositronAlliance:2019cpi}
{\scshape Electron-Positron Alliance} collaboration, A.~Badea, A.~Baty,
  P.~Chang, G.~M. Innocenti, M.~Maggi, C.~Mcginn et~al., \emph{{Measurements of
  two-particle correlations in $e^+e^-$ collisions at 91 GeV with ALEPH
  archived data}},
  \href{https://doi.org/10.1103/PhysRevLett.123.212002}{\emph{Phys. Rev. Lett.}
  {\bfseries 123} (2019) 212002},
  [\href{https://arxiv.org/abs/1906.00489}{{\ttfamily 1906.00489}}].

\bibitem{Chen:2021uws}
Y.~Chen et~al., \emph{{Jet energy spectrum and substructure in $e^+e^-$
  collisions at 91.2 GeV with ALEPH Archived Data}},
  \href{https://doi.org/10.1007/JHEP06(2022)008}{\emph{JHEP} {\bfseries 06}
  (2022) 008}, [\href{https://arxiv.org/abs/2111.09914}{{\ttfamily
  2111.09914}}].

\bibitem{Chen:2023njr}
Y.-C. Chen et~al., \emph{{Long-range near-side correlation in $e^+e^-$
  collisions at 183-209 GeV with ALEPH archived data}},
  \href{https://doi.org/10.1016/j.physletb.2024.138957}{\emph{Phys. Lett. B}
  {\bfseries 856} (2024) 138957},
  [\href{https://arxiv.org/abs/2312.05084}{{\ttfamily 2312.05084}}].

\bibitem{Electron-PositronAlliance:2025fhk}
{\scshape Electron-Positron Alliance} collaboration, H.~Bossi et~al.,
  \emph{{Energy Correlators from Partons to Hadrons: Unveiling the Dynamics of
  the Strong Interactions with Archival ALEPH Data}},
  \href{https://arxiv.org/abs/2511.00149}{{\ttfamily 2511.00149}}.

\bibitem{Electron-PositronAlliance:2025hze}
{\scshape Electron-Positron Alliance} collaboration, A.~Badea et~al.,
  \emph{{Unbinned measurement of thrust in $e^+e^-$ collisions at $\sqrt{s}$ =
  91.2 GeV with ALEPH archived data}},
  \href{https://arxiv.org/abs/2510.22038}{{\ttfamily 2510.22038}}.

\bibitem{DELPHI:2024opendata}
{\scshape DELPHI} collaboration, DELPHI, ``{DELPHI Collaboration releases its
  entire data collection}.'' CERN Open Data Portal, 8, 2024.

\bibitem{Zhang:2026esd}
J.~Zhang, Y.~Chen, K.~Lee, I.~Moult, C.~Baldenegro, H.~Bossi et~al.,
  \emph{{Analysis note: one-point charge correlator with DELPHI Open Data}},
  \href{https://arxiv.org/abs/2608.15367}{{\ttfamily 2608.15367}}.

\bibitem{DELPHI:1990cdc}
{DELPHI Collaboration}, \emph{{The DELPHI detector at LEP}},
  \href{https://doi.org/10.1016/0168-9002(91)90282-U}{\emph{Nucl. Instrum.
  Meth. A} {\bfseries 303} (1991) 233--276}.

\bibitem{DELPHI:1995dsm}
{\scshape DELPHI} collaboration, P.~Abreu et~al., \emph{{Performance of the
  DELPHI detector}},
  \href{https://doi.org/10.1016/0168-9002(96)00463-9}{\emph{Nucl. Instrum.
  Meth. A} {\bfseries 378} (1996) 57--100}.

\bibitem{DELPHI:OpenData:short94_c2}
{DELPHI Collaboration}, \emph{{Hadronic} $z$ {decay data sample at}
  $\sqrt{s}=91.25$~{GeV from 1994 LEP-1 run (short DST, processing tag c2)}},
  2024.
\newblock 10.7483/OPENDATA.DELPHI.0XNE.G96F.

\bibitem{DELPHI:OpenData:short95_d2}
{DELPHI Collaboration}, \emph{{Hadronic} $z$ {decay data sample at}
  $\sqrt{s}=91.25$~{GeV from 1995 LEP-1 run (short DST, processing tag d2)}},
  2024.
\newblock 10.7483/OPENDATA.DELPHI.K4LR.4PQ4.

\bibitem{DELPHI:2024policy}
{\scshape DELPHI} collaboration, DELPHI, ``{DELPHI data preservation, re-use,
  and open access policy}.'' CERN Open Data Portal, 8, 2024.
\newblock 10.7483/OPENDATA.DELPHI.0X8O.FBUH.

\bibitem{Bierlich:2022pfr}
C.~Bierlich et~al., \emph{{A comprehensive guide to the physics and usage of
  PYTHIA 8.3}},
  \href{https://doi.org/10.21468/SciPostPhysCodeb.8}{\emph{SciPost Phys.
  Codeb.} {\bfseries 2022} (2022) 8},
  [\href{https://arxiv.org/abs/2203.11601}{{\ttfamily 2203.11601}}].

\bibitem{Skands:2014pea}
P.~Skands, S.~Carrazza and J.~Rojo, \emph{{Tuning PYTHIA 8.1: the Monash 2013
  Tune}}, \href{https://doi.org/10.1140/epjc/s10052-014-3024-y}{\emph{Eur.
  Phys. J. C} {\bfseries 74} (2014) 3024},
  [\href{https://arxiv.org/abs/1404.5630}{{\ttfamily 1404.5630}}].

\bibitem{DELSIM}
{DELPHI Collaboration}, \emph{{DELSIM, DELPHI Event Generation and Detector
  Simulation User's Guide}},  Tech. Rep. DELPHI 89-68 PROG 143, CERN, 1989.

\bibitem{DELPHI:OpenData:kk2f_pythia_94}
{DELPHI Collaboration}, \emph{{KK2f + PYTHIA Monte Carlo sample of} $e^+e^- \to
  q\bar{q}$ {at} $\sqrt{s}=91.25$~{GeV, 1994 detector configuration}},  2024.
\newblock 10.7483/OPENDATA.DELPHI.P4BN.GFID.

\bibitem{DELPHI:OpenData:kk2f_pythia_95}
{DELPHI Collaboration}, \emph{{KK2f + PYTHIA Monte Carlo sample of} $e^+e^- \to
  q\bar{q}$ {at} $\sqrt{s}=91.25$~{GeV, 1995 detector configuration}},  2024.
\newblock 10.7483/OPENDATA.DELPHI.OJDR.XYNG.

\bibitem{Jadach:1999vf}
S.~Jadach, B.~Ward and Z.~Was, \emph{{The Precision Monte Carlo event generator
  KK for two fermion final states in e+ e- collisions}},
  \href{https://doi.org/10.1016/S0010-4655(99)00402-2}{\emph{Comput. Phys.
  Commun.} {\bfseries 130} (2000) 260--325},
  [\href{https://arxiv.org/abs/hep-ph/9912214}{{\ttfamily hep-ph/9912214}}].

\bibitem{Sjostrand:2000wi}
T.~Sjostrand, P.~Eden, C.~Friberg, L.~Lonnblad, G.~Miu, S.~Mrenna et~al.,
  \emph{{High-energy physics event generation with PYTHIA 6.1}},
  \href{https://doi.org/10.1016/S0010-4655(00)00236-8}{\emph{Comput. Phys.
  Commun.} {\bfseries 135} (2001) 238--259},
  [\href{https://arxiv.org/abs/hep-ph/0010017}{{\ttfamily hep-ph/0010017}}].

\bibitem{DELPHI:2003yqh}
{DELPHI Collaboration}, \emph{{The measurement of $\alpha_s$ from event shapes
  with the DELPHI detector at the highest LEP energies}},
  \href{https://doi.org/10.1140/epjc/s2003-01530-0}{\emph{Eur. Phys. J. C}
  {\bfseries 37} (2004) 1--23},
  [\href{https://arxiv.org/abs/hep-ex/0311019}{{\ttfamily hep-ex/0311019}}].

\bibitem{Zhang:2025delphiEEC}
J.~Zhang et~al., \emph{{Measurement of thrust and track energy-energy
  correlator in $e^{+}e^{-}$ collisions at 91.2 GeV with DELPHI open data}},
  \href{https://arxiv.org/abs/2510.18762}{{\ttfamily 2510.18762}}.

\bibitem{hungarianMatching}
H.~W. Kuhn, \emph{The hungarian method for the assignment problem},
  \href{https://doi.org/10.1002/nav.3800020109}{\emph{Naval Research Logistics
  Quarterly} {\bfseries 2} (1955) 83--97}.

\bibitem{SLD:1996gjt}
{\scshape SLD} collaboration, K.~Abe et~al., \emph{{First measurement of the
  left-right charge asymmetry in hadronic Z boson decays and a new
  determination of sin**2 theta(W)(eff)}},
  \href{https://doi.org/10.1103/PhysRevLett.78.17}{\emph{Phys. Rev. Lett.}
  {\bfseries 78} (1997) 17--21},
  [\href{https://arxiv.org/abs/hep-ex/9609019}{{\ttfamily hep-ex/9609019}}].

\bibitem{Elsing:2000fv}
M.~Elsing, \emph{{The DELPHI silicon tracker in the global pattern
  recognition}},
  \href{https://doi.org/10.1016/S0168-9002(00)00175-3}{\emph{Nucl. Instrum.
  Meth. A} {\bfseries 447} (2000) 76--89},
  [\href{https://arxiv.org/abs/hep-ex/0001064}{{\ttfamily hep-ex/0001064}}].

\bibitem{Osterberg:1998xxx}
K.~Osterberg, \emph{Ph.D. thesis}, Ph.D. thesis, Helsinki, University of
  Helsinki, 1998.

\bibitem{DELPHI:2000uri}
{DELPHI Collaboration}, \emph{{Consistent measurements of $\alpha_s$ from
  precise oriented event shape distributions}},
  \href{https://doi.org/10.1007/s100520000421}{\emph{Eur. Phys. J. C}
  {\bfseries 14} (2000) 557--584},
  [\href{https://arxiv.org/abs/hep-ex/0002026}{{\ttfamily hep-ex/0002026}}].

\bibitem{Lee:2026bcq}
K.~Lee and I.~Moult, \emph{{Flavor Flux Correlators, Forward-Backward
  Asymmetries, and Anomalies}},
  \href{https://arxiv.org/abs/2609.13366}{{\ttfamily 2609.13366}}.

\bibitem{FCC:2018byv}
{\scshape FCC} collaboration, M.~Benedikt et~al., \emph{{FCC-ee: The Lepton
  Collider}}, \href{https://doi.org/10.1140/epjst/e2019-900045-4}{\emph{Eur.
  Phys. J. ST} {\bfseries 228} (2019) 261--623}.

\bibitem{CEPCStudyGroup:2018ghi}
{\scshape CEPC Study Group} collaboration, M.~Dong et~al., \emph{{CEPC
  Conceptual Design Report: Volume 2 - Physics \& Detector}},
  \href{https://arxiv.org/abs/1811.10545}{{\ttfamily 1811.10545}}.

\end{thebibliography}\endgroup

\end{document}